\documentclass[12pt]{JHEP}
\usepackage{amsmath,epsfig}
\usepackage{amssymb,amsfonts}
\usepackage{latexsym}
\usepackage{epsfig}
\newbox\pippobox
\def\be{\begin{equation}}
\def\ee{\end{equation}}
\def\bea{\begin{eqnarray}}
\def\eea{\end{eqnarray}}

\def\6{\partial}

\def\a{\alpha}

\def\C0{{\bf C_0}}
\def\Y0{{\bf Y_0}}
\def\G0{{\bf G_0}}

\def\sq
\def\a{\alpha}

\title{Thermodynamics of deformed AdS$_5$ model with a positive/negative
quadratic correction in graviton-dilaton system
}
\author{Danning Li $^{1}$, Song He $^{1,2}$, Mei Huang$^{1,3}$,
Qi-Shu Yan $^{2}$ \\
$^{1}$ {Institute of High Energy Physics, Chinese Academy of
Sciences, Beijing, China} \\
$^{2}$ {College of Physical Sciences, Graduate University of Chinese
Academy of Sciences, Beijing, China}\\
$^{3}$ {Theoretical Physics Center for Science Facilities, Chinese
Academy of Sciences, Beijing, China} \\
}

\date{\today}

\abstract{By solving the Einstein equations of the graviton coupling
with a real scalar dilaton field, we establish a general framework
to self-consistently solve the geometric background with black-hole
for any given phenomenological holographic models. In this framwork,
we solve the black-hole background, the corresponding dilaon field
and the dilaton potential for the deformed AdS$_5$ model with a
positive/negative quadratic correction. We systematically
investigate the thermodynamical properties of the deformed AdS$_5$
model with a positive and negative quadratic correction,
respectively, and compare with lattice QCD on the results of the
equation of state, the heavy quark potential, the Polyakov loop and
the spatial Wilson loop. We find that the bulk thermodynamical
properties are not sensitive to the sign of the quadratic
correction, and the results of both deformed holographic QCD models
agree well with lattice QCD result for pure SU(3) gauge theory.
However, the results from loop operators favor a positive quadratic
correction, which agree well with lattice QCD result. Especially,
the result from the Polyakov loop excludes the model with a negative
quadratic correction in the warp factor of ${\rm AdS}_5$.
}


\begin{document}

\maketitle


\section{Introduction}
In recent decade, the discovery of the anti-de Sitter/conformal
field theory (AdS/CFT) correspondence and the conjecture of the
gravity/gauge duality \cite{dual} has been widely used to understand
strongly coupled dense and hot quark matter
\cite{Kovtun:2004de,etas-adscft,densematter-adscft,
Gubser-T,Gursoy-T1,Gursoy-T2,Pirner-T,Andreev-T,Andreev-T1,Andreev-T2,Andreev-T3,Noronha-T,Kajantie-T}.

One of the most successful examples of using thermal ${\cal N}= 4$
super-Yang-Mills theory (SYM) in quantum chromodynamics (QCD) at
finite temperature is the small value of shear viscosity over
entropy density $\eta/s= 1/4\pi$ \cite{Kovtun:2004de}, which is very
close to the value used to fit the RHIC data of elliptic flow $v_2$
\cite{Hydro, Hydro-Teaney}. It is now believed that the system
created at RHIC is a strongly coupled quark-gluon plasma (sQGP) and
behaves like a nearly "perfect" fluid \cite{RHIC-EXP,RHIC-THEO}.
However, lattice QCD results show that the bulk viscosity over
entropy density ratio $\zeta/s$ rises dramatically up to the order
of $1.0$ near the critical temperature $T_c$
\cite{LAT-xis-KT,LAT-xis-KKT,LAT-xis-Meyer}. The sharp peak of bulk
viscosity at $T_c$ has also been observed in QCD effective models
\cite{bulk-Paech-Pratt,Li-Mao-Huang,bulk-Chen}. The large bulk
viscosity near phase transition is related to the non-conformal
equation of state \cite{LAT-EOS-G, LAT-EOS-Nf2}, e.g, the square of
sound velocity is around 0.07, which is much smaller than the
conformal value $1/3$. In the SYM plasma, the square of sound
velocity is always $1/3$ and the bulk viscosity $\zeta$ always
vanishes at all temperatures.

These facts demonstrate that strongly coupled quark gluon plasma
near $T_c$ is not conformal invariant. In order to mimic the QCD
equation of state, much effort has been put to find the gravity dual
of gauge theories which break the conformal symmetry, e.g,
\cite{Gubser-T,Gursoy-T1,Gursoy-T2,Pirner-T}, where a real scalar
dilaton field background has been introduced to couple with the
graviton. Refs.\cite{Gubser-T} and \cite{Gursoy-T1,Gursoy-T2} have
used different dilaton potentials as input, while
Ref.\cite{Pirner-T} has used QCD $\beta$-function as input.

The AdS/CFT approach has also been widely applied to describe
non-perturbative phenomenology in the vacuum, where conformal
symmetry is also needed to be broken
\cite{EKSS2005,TB:05,DaRold2005,D3-D7,D4-D6,SS,Dp-Dq,Karch:2006pv,Andreev:2006ct,
Zuo:2009dz,deTeramond:2009xk,Shock-2006,Ghoroku-Tachibana,Csaki:2006ji,
Gursoy,Zeng:2008sx,Pirner:2009gr,He:2010ye}. The most economic way
of breaking the conformal symmetry is to add a proper deformed warp
factor in front of the ${\rm AdS}_5$ metric structure, which can
capture the main features of non-perturbative QCD physics. For
example, Andreev-Zakharov proposed a positive quadratic correction,
$e^{cz^2}$ with $z$ the fifth dimension coordinate and $c>0$, in the
deformed warp factor of ${\rm AdS}_5$ geometry, which can help to
realize the linear heavy quark potential \cite{Andreev:2006ct}. The
linear heavy quark potential can also be obtained by introducing
other deformed warp factors, e.g. the deformed warp factor which
mimics the QCD running coupling \cite{Pirner:2009gr}, and the
logarithmic correction with an explicit IR cutoff $\log
\frac{z_{IR}-z}{z_{IR}}$ \cite{He:2010ye}. To produce the linear
Regge behavior of the hadron excitations, Karch-Katz-Son-Stephanov
\cite{Karch:2006pv} proposed the soft-wall ${\rm AdS}_5$ model or
KKSS model by introducing a quadratic dilaton background in the 5D
meson action, whose effect in some sense looks like introducing a
negative quadratic correction, $e^{-cz^2}$, in the warp factor of
the ${\rm AdS}_5$ geometry. However, it is worthy of mentioning that
the model with a quadratic correction in the warp factor of the
metric is not equivalent to the model with a quadratic correction in
the dilaton background. A positive quadratic correction $e^{cz^2}$
in the dilaton background of the 5D hadron action has also been used
to investigate hadron spectra \cite{Zuo:2009dz,deTeramond:2009xk},
however, higher spin excitations in this background will get
imaginary mass \cite{Dp-Dq,KKSS-2}.

From the above studies of constructing the holographic QCD (hQCD)
models for describing the heavy quark potential and the light hadron
spectra, we have observed that a quadratic background correction is
related to the confinement property, i.e. the linear quark
anti-quark potential and the linear Regge behavior. Furthermore, we
have observed that to produce the linear heavy quark potential, the
positive quadratic correction has been introduced in the deformed
warp factor, while to produce the linear Regge behavior, the
negative quadratic correction is introduced in the dilaton
background in the 5D hadron action.

It is natural to ask the connection between the quadratic correction
in the metric and dilaton background, and which sign is consistent
with real QCD data. If we assume that there is no direct connection
between the dilaton field and metric background, these two types of
models cannot be contrasted. However, the metric structure of the
hQCD model and its corresponding dilaton background can be
self-consistently solved from the Einstein equation in the
non-critical string framework or the 5D Einstein dilaton system,
e.g, as described in Ref. \cite{Gursoy}. In this framework, the
connection between the metric structure and the dilaton background
can be established.

In Ref.\cite{He:2010ye}, we have numerically solved the dilaton
background at zero temperature for the deformed ${\rm AdS}_5$ model
with positive and negative quadratic correction, respectively. In
this paper, we extend our previous study in the non-critical string
framework at zero temperature to finite temperature, and
self-consistently solve the black-hole background, dilaton field and
the dilaton potential. In order to further pin down the sign of the
quadratic correction in the hQCD model, we systematically
investigate the thermodynamic properties of these two deformed ${\rm
AdS}_5$ models, including the equation of state, the heavy quark
potential, the spatial string tension, and the Polyakov loop at
finite temperature. By comparing with lattice results, we find that
the equation of state is not sensitive to the sign of the quadratic
correction. However, the results from loop operators favor a
positive quadratic correction. Especially, the result from the
Polyakov loop excludes the model with negative quadratic correction
in the warp factor of ${\rm AdS}_5$.

The paper is organized as follows. In
Sec.\ref{section-graviton-dilaton}, we derive the general formulae
for equations of motion in classical 5D gravity-dilaton system, and
solve the black-hole background, the dilaton field and its
corresponding dilaton potential for any given hQCD model. In
Sec.\ref{sec-solution-hQCD}, we self-consistently solve the dual
black-hole background, the dilaton field and its dilaton potential
for the hQCD models with positive and negative quadratic correction
in the deformed warp factor of ${\rm AdS}_5$ metric background. In
Sec.\ref{section-eos}, we study the phase transition and the
equation of state, including the entropy density, the pressure
density, the energy density and the square of sound velocity for the
two hQCD models with quadratic corrections. In
Sec.\ref{section-loop}, we calculate the heavy quark potential, the
Polyakov loop and the spatial Wilson loop at finite temperature and
compare our results with lattice data. The summary and discussion is
given in Sec.\ref{section-summary}. We also list several special
exact solutions of the general graviton-dilaton system in Appendix
\ref{appendix-solution}, and the method of extracting the spatial
string tension in Appendix \ref{appendix-spatialtension}.

\section{5D Einstein dilaton system with black-hole background}
\label{section-graviton-dilaton}

\subsection{Minimal non-critical string framework}

For low energy IIA/IIB supergravity, the classical action in the
string frame has the following form \cite{string-book}:
\begin{equation} \label{supergravity}
S_{IIA/IIB}=S_{NS}+S_{R}+S_{CS} + S_{fermion}.
\end{equation}
Where $S_{fermion}$ is the action from the fermionic part, and
$S_{NS},S_{R},S_{CS}$ are bosonic part actions for Neveu-Schwarz
(NS) sector of graviton-dilton coupling, Ramond (R) kinetic term for
the form field, Chern-Simons term, respectively. The NSNS sector
action takes the form of
\begin{equation}
S_{NS}=\frac{1}{2k_{10}^2}\int d^{10}\sqrt{-G_s}
e^{-2\phi}\left(R+4\partial_\mu \phi
\partial^\mu \phi-\frac{1}{2} \left|H_3\right|^2\right).
\end{equation}
The RR part action and Chern-Simons term take different forms for
Type-IIA and IIB supergravity, respectively:
\begin{eqnarray}
S_{R}|_{IIA}&=&\frac{1}{2k_{10}^2}\int d^{10}\sqrt{-G_s}
\left(\left|F_2\right|^2+
 \left|\widetilde{F_4}\right|^2 \right), \nonumber\\
 S_{CS}|_{IIA}&=& \frac{1}{2k_{10}^2}\int B_2\wedge F_4\wedge F_4,
\nonumber \\
S_{R}|_{IIB}&=&\frac{1}{2k_{10}^2}\int d^{10}\sqrt{-G_s}
\left(\left|F_1\right|^2+
\frac{1}{2} \left|\widetilde{F_3}\right|^2+\left|\widetilde{F_5}\right|^2\right), \nonumber\\
S_{CS}|_{IIB}&=& \frac{1}{2k_{10}^2}\int C_4\wedge H_3\wedge F_3.
\end{eqnarray}
Where $k_{10}$ the 10-dimension Newton coupling constant, $G_s$ the
metric in the string frame, $R, \phi$ are scalar curvature and the
dilaton field, respectively. $B$ is anti-symmetric 2-form tensor
field, $C$ is R-R form field, $H,F$ are field strengths of $B,C$,
and $F=e^{-\phi}\tilde{F}$. $F_2,F_4$ only appear in IIA and
$F_1,F_3,F_5$ only appear in IIB supergravity.

We are interested in building a 5-dimensional effective gravity dual
to a 4-dimensional non-Abelian gauge theory. We follow the argument
in Ref.\cite{Gursoy} to derive the minimal 5D effective
graviton-dilaton system from 10D supergravity
Eq.(\ref{supergravity}).

1) For pure Yang-Mills gauge theory, there is no fermionic degrees
of freedom, therefore, we neglect the $S_{fermion}$ in
Eq.(\ref{supergravity}) for simplicity.

2) In the NSNS sector, one should note that we ignore the effect of
$H_3=dB_2$, which is normally taken to be zero because it induces
noncommutative effect in the string theory. This leads to the
vanishing of the Chern-Simons action, i.e. $S_{CS}=0$.

3) The RR part includes six fields strengths
$F_0,F_1,F_2,F_3,F_4,F_5$, where only $F_0,F_1,F_2$ are independent
due to the Poincare dual relationship. $F_0$ couples to bulk
instantons, and its dual $F_5$ $F_0$ generates a four-form field
stems from $D_3$ branes related to the $U(N_c)$ gauge group. $F_1=d
C_0$ originates from axion field $C_0=a$ which is dual to
$Tr[F\wedge F]$ in the gauge theory. $F_4\sim F_1$ and its
associated three-form $C_3$ couples to domain walls. $F_2$ generates
a vector and couples minimally to the baryon density, and it breaks
Lorentz invariance in the QCD vacuum. Therefore, $F_2$ and its dual
$F_3$ can be safely neglected.

From the above analysis, the relevant low-lying fields, which are
expected to be dual to the lowest dimension operators in the gauge
theory, are the graviton $g_{\mu\nu}$, the dilaton $\phi$, and the
RR axion $a$. One can put the solution of $F_5$ into
Eq.(\ref{supergravity}) and then integrate out the inner five
compact space-time $M_5$. Finally, one can obtain the minimal
non-critical 5D effective gravity action as following:
\begin{equation} \label{minimal-String-action}
S_{5D}=\frac{1}{16 \pi G_5}\int d^5 x \sqrt{-g^S} e^{-2 \phi}
\left(R + 4\partial_\mu \phi
\partial^\mu \phi- V_S(\phi)\right)
\end{equation}
Where $G_5$ is 5D Newton constant, $g^S$ and $V_S(\phi)$ are the 5D
metric and dilaton potential in the string frame, respectively.

In the minimal non-critical string framework, the metric structure,
the dilaton field and the dilaton potential should be solved
self-consistently from the Einstein equations. One can use the
dilton potential as input to solve the metric structure and the
dilaton field. There are several different choices for the dilaton
potential from several groups. The scalar field or dilaton field
$\phi$ encodes the running of the Yang-Mills gauge theory's coupling
$\lambda$. In the framework of G{\"u}rsoy -Kiritsis-Nitti (GKN)
\cite{Gursoy, Gursoy-T1,Gursoy-T2}, the renormalized dilaton field
$\phi$ has been defined as $\lambda=e^{\phi}$, and the dilaton
potential can be solved from the input of QCD running coupling
function, i.e, $\beta(\lambda)$-function. In Ref. \cite{Gubser-T},
in order to mimic the QCD equation of state, the authors chooses the
dilaton potential as $V(\phi)=\frac{-12{\rm cosh} \gamma\phi
+b\phi^2}{L^2}$ where $L$ is the radius of the ${\rm AdS}_5$.

However, in the graviton-dilaton coupling system, for the given
dilaton potential in \cite{Gubser-T,Gursoy-T1,Gursoy-T2,Gursoy}, the
solution of the metric structure is very complicated. Moreover, for
finite temperature case, the temperature dependence of the dilaton
potential is not obvious. In Ref.\cite{He:2010ye}, instead of input
the dilaton potential, we solve the diltaon potential for given
metric structure of deformed ${\rm AdS}_5$ model, which can capture
some features of QCD phenomenology at zero temperature. In this
work, we will extend the minimal non-critical string framework to
finite temperature, and investigate the phase transitions, the
equation of state and the temperature dependence of loop operators.

\subsection{Graviton-dilaton system with dual black-hole for hQCD model}

The finite temperature dynamics of gauge theories, has a natural
holographic counterpart in the thermodynamics of black-holes on the
gravity side. Our holographic model is defined in the string frame,
it is convenient to calculate the vacuum expectation value of the
loop operator in the string frame, however it is more convenient to
work out the gravity solution and study equation of state in the
Einstein frame. Therefore, for later use, we will firstly derive the
relation between the string frame and the Einstein frame.


If the metric in the Einstein frame $g^E_{\mu\nu}$ and its
corresponding metric structure in the string frame $g^S_{\mu\nu}$
are connected by the scaling transformation
\begin{equation}
g^S_{\mu\nu} = e^{-2\Omega}g^E_{\mu\nu},
\end{equation}
then the scalar curvature in the Einstein frame and string frame has
the following exact relationship:
\begin{eqnarray}
e^{-2 \Omega} R^S &=& R^E -(D-1)(D-2) \partial_\mu \Omega
\partial^\mu \Omega+ 2(D-1)\nabla^2 \Omega,
\end{eqnarray}
with $D$ the dimension, and $\nabla^2$ is defined by
$\frac{1}{\sqrt{-g^E}}
\partial_\mu \sqrt{-g^E}
\partial^\mu$, which is useful to derive the exact relation
between two actions in string frame or Einstein frame.

In the case of $D=5$, we have
\begin{eqnarray}
& & \int \sqrt{-g^S} e^{m \Omega} R^S = \int \sqrt{-g^E}
      e^{(m-3)\Omega}
      \left[ R^E -12 \partial_\mu \Omega \partial^\mu\Omega
      + 8(m-3) \nabla^2 \Omega \right], \\
& & \int \sqrt{-g^S} e^{m\Omega}\left( g_S^{\mu\nu} \partial_\mu
\Omega
\partial_\nu \Omega\right) = \int \sqrt{-g^E} e^{(m-5)\Omega}\left(
e^{2\Omega}g_E^{\mu\nu} \partial_\mu \Omega \partial_\nu
\Omega\right).
\end{eqnarray}
The dilaton potentials in two different frames own the following
relationship
\begin{eqnarray}
V_S(\Omega)=V_E(\Omega) e^{2\Omega}.
\end{eqnarray}
By setting $m=3$ and $\Omega=-\frac{2}{3}\phi$, we have
\begin{equation}
V_S=V_E e^{\frac{-4\phi}{3}},
\end{equation}
and the following relation:
\begin{eqnarray}\label{string-enssteinframe}
& & \int \sqrt{-g^S} e^{-2 \phi} \left(R^S + 4\partial_\mu \phi
\partial^\mu \phi- V_S(\phi)\right) \nonumber \\
&=&\int \sqrt{-g^E} \left[ R^E
-\frac{4}{3}
\partial_\mu \phi \partial^\mu \phi - V_E(\phi)\right].
\end{eqnarray}
Therefore, Eq.(\ref{minimal-String-action}) in the Einstein frame
becomes:
\begin{equation} \label{minimal-Einstein-action}
S_{5D}=\frac{1}{16 \pi G_5} \int d^5 x
\sqrt{-g^E}\left(R-\frac{4}{3}\partial_{\mu}\phi\partial^{\mu}\phi-V_E(\phi)\right)
\end{equation}


Adding the black-hole background to the holographic QCD model
constructed from vacuum properties, in the string frame we have
\begin{equation} \label{metric-stringframe}
ds_S^2=\frac{L^2
e^{2A_s}}{z^2}\left(-f(z)dt^2+\frac{dz^2}{f(z)}+dx^{i}dx^{i}\right),
\end{equation}
with $L$ the radius of ${\rm AdS}_5$. The metric in the string frame
is useful to calculate the loop operator. To derive the Einstein
equations and to study the thermodynamical properties of equation of
state, we transform it to the Einstein frame
\begin{eqnarray} \label{metric-Einsteinframe}
ds_E^2=\frac{L^2 e^{2A_s-\frac{4\phi}{3}}}{z^2}\left(-f(z)dt^2
+\frac{dz^2}{f(z)}+dx^{i}dx^{i}\right).
\end{eqnarray}

The general Einstein equations from the action
(\ref{minimal-Einstein-action}) takes the form of:
\begin{eqnarray} \label{EOM}
E_{\mu\nu}+\frac{1}{2}g^E_{\mu\nu}\left(\frac{4}{3}
\partial_{\mu}\phi\partial^{\mu}\phi+V_E(\phi)\right)
-\frac{4}{3}\partial_{\mu}\phi\partial_{\nu}\phi=0.
\end{eqnarray}
By using Eqs.(\ref{metric-Einsteinframe}) and
(\ref{minimal-Einstein-action}), we can derive the following
nontrivial Einstein equations in $(t,t), (z,z)$ and $(x_1, x_1)$
spaces, respectively:
\begin{eqnarray} \label{tt}
& & A_s''(z)+A_s'(z)\left(\frac{ f'(z)}{2 f(z)}-\frac{4}{3} \phi
'(z)-\frac{2}{z}+A_s'(z)\right)-f'(z)\left(\frac{ \phi '(z)}{3 f(z)}
+\frac{1}{2 z f(z)}\right) \nonumber\\
& & -\frac{2 \phi ''(z)}{3}+\frac{2}{3}\phi '(z)\left( \phi
'(z)+\frac{2 }{z}\right) +\frac{L^2  V_E(\phi (z))}{6 z^2 f(z)}e^{2
A_s(z)-\frac{4 \phi (z)}{3}}+\frac{2}{z^2}=0\end{eqnarray}

\begin{eqnarray} \label{rr}
\phi '(z)^2&-&\phi '(z)\left(4 A_s'(z) +\frac{f'(z) }{2
f(z)}-\frac{4 }{z}\right)
+A_s'(z)\left(\frac{3 f'(z)}{4 f(z)}-\frac{6 }{z}+3 A_s'(z)\right)\nonumber\\
&-&\frac{3 f'(z)}{4 r f(z)}+\frac{3}{z^2}+\frac{L^2  V_E(\phi
(z))}{4 z^2 f(z)}e^{2 A_s(z)-\frac{4 \phi (z)}{3}}=0 \end{eqnarray}

\begin{eqnarray} \label{xx}
&{}&f''(z)+ \left(6 A_s'(z)-4\text{  }\phi '(z)-\frac{6
}{z}\right)f'(z) +\frac{L^2 V_E(\phi (z))}{z^2}e^{2 A_s(z)-\frac{4
\phi (z)}{3}} \nonumber\\
& & +f(z) (6 A_s''(z)-\frac{4 A_s'(z) \left(2 z \phi
'(z)+3\right)}{z} +6 A_s'(z)^2 \nonumber \\
& & + \frac{12}{z^2}-4 \phi ''(z)+4 \phi '(z)^2+\frac{8 \phi
'(z)}{z})=0
\end{eqnarray}

One should notice that the above three equations are not
independent. There are only two independent functions, one left
equation is to check the consistence of the solutions. To simplify
the above equations, one can use the following two equations without
dilaton potential $V_E$:
\begin{eqnarray}
\phi ''(z)-\left(2 A_s'(z) -\frac{2 }{z}\right)\phi '(z)
-\frac{3 A_s''(z)}{2}+\frac{3}{2} A_s'(z)^2-\frac{3 A_s'(z)}{z}&=&0, \nonumber\\
f''(z)+\left(3 A_s'(z) -2 \phi '(z)-\frac{3 }{z}\right)f'(z)&=&0.
\end{eqnarray}
The EOM of the dilaton field is given as following
\begin{equation}
\label{fundilaton} \frac{8}{3} \partial_z
\left(\frac{L^3e^{3A_s(z)-2\phi} f(z)}{z^3}
\partial_z \phi\right)-
\frac{L^5e^{5A_s(z)-\frac{10}{3}\phi}}{z^5}\partial_\phi V_E=0.
\end{equation}

For any holographic QCD model with given metric structure $A_s(z)$
in the string frame, we can derive the general solutions to the
Einstein equations, which take the following form:
\begin{eqnarray} \label{general-solution}
 \phi(z)&=&\phi _ 0 + \phi _ 1\int_ 0^
 z\frac {e^{2 A_s(x)}} {x^2}\, dx + \frac {3 A_s(z)} {2} \nonumber \\
  &+ & \frac{3}{2}\int_ 0^
    z\frac {e^{2 A_s(x)}\int _ 0^{x}y^2 e^{-2 A_s(y)} A_s'(y) {}^2 dy} {x^2}\, dx,\nonumber\\
f(z)&=&f_0+f_ 1\left (\int_ 0^z x^3 e^{2\phi (x) - 3 A_s (x)}\,
dx\right), \nonumber\\
V_E(\phi)&=&\frac {e^{\frac {4\phi (z)}{3} - 2
A_s (z)}}{L^2} \nonumber \\
   & & \left (z^2 f^{''}(z) - 4 f (z)\left (3 z^2 A_s''(z) - 2
z^2\phi^{''}(z) + z^2\phi' (z)^2 + 3 \right) \right).
\end{eqnarray}
Where $\phi_0,\phi_1,f_0,f_1$ are constants of integration. In
Appendix \ref{appendix-solution}, we list several simple exact
solutions of the Einstein equations by using
Eq.(\ref{general-solution}).

\section{Dual black-hole solution for the hQCD model with quadratic correction}
\label{sec-solution-hQCD}

From the experiences of constructing holographic QCD models for
describing the heavy quark potential and the light hadron spectra,
we have learnt that a quadratic background correction is related to
the confinement property, i.e. the linear quark anti-quark potential
and the linear Regge behavior. A positive quadratic correction,
$e^{cz^2}$ with $c>0$, in the deformed warp factor of ${\rm AdS}_5$
can help to realize the linear heavy quark potential
\cite{Andreev:2006ct}. A quadratic dilaton background in the 5D
meson action, whose effect in some sense looks like introducing a
negative quadratic correction, $e^{-cz^2}$, in the warp factor of
the ${\rm AdS}_5$ geometry, is helpful to realize the linear Regge
behavior of hadron excitations \cite{Karch:2006pv}.

Therefore, we introduce the following holographic QCD model with
positive and negative quadratic correction in the deformed warp
factor of ${\rm AdS}_5$ in Eq.(\ref{metric-stringframe}), i.e. we
take
\begin{equation} \label{ourmodel}
A_s(z)= c k^2 z^2
\end{equation}
with $c=\pm$ indicating the positive and negative quadratic
correction, respectively. The main purpose in this section is to
solve the dual black-hole background, corresponding dilaton field,
and dilaton potential in the Graviton-dilaton system for the given
hQCD model Eq.(\ref{ourmodel}).  To get the solution of equations
Eq.(\ref{tt})-Eq.(\ref{fundilaton}), we impose the asymptotic
$AdS_5$ condition $f(0)=1$ near the UV boundary $z\sim 0$,  and
require $\phi, f$ to be finite $z=0, z_h$ with $z_h$ the black-hole
horizon. Fortunately, we find that the solution of the black-hole
background takes the form of,
\begin{eqnarray} \label{solu-f}
f(z)= 1- f_{c}^h \int_0^{k z} x^3 \exp \left(\frac{3}{2} c x^2
\left( H_c(x/k)-1 \right)\right) dx,
\end{eqnarray}
with
\begin{eqnarray}\label{fc}
f_{c}^h= \frac{1}{\int_0^{k z_h} x^3 \exp \left(\frac{3}{2} c x^2
\left( H_c(x/k)-1 \right)\right)dx }, \label{fc-zh}
\end{eqnarray}
and
\begin{equation} \label{Hc}
H_{c}(z)=\, _2F_2\left(1,1;2,\frac{5}{2};2 c k^2 z^2\right).
\end{equation}

It is worthy of mentioning that in
Ref.\cite{Andreev-T1,Andreev-T2,Andreev-T3}, the authors also
investigated the thermodynamic properties of the deformed ${\rm
AdS}_5$ model with positive quadratic correction in the warp factor,
however, there the black-hole background was not solved
self-consistently from the graviton-dilaton system, but just taken
from the ${\rm AdS}_5$ Schwarz blak-hole (AdS-SW BH) background
\begin{equation}
\label{pureAdSBH} f(z)=1-z^4/z_T^4,
\end{equation}
with $z_T=1/\pi T$.

The solutions for the dilaton field and the dilaton potential take
the forms as following:
\begin{eqnarray} \label{solu-phi}
\phi(z) &=& \frac{3}{4} c k^2 z^2(1+ H_c(z)),
\end{eqnarray}

\begin{eqnarray} \label{solu-V-c}
V_{E}^{c}(z)&=&\frac{3 e^{c k^2 z^2 \left(-1+H_c(z)\right)}}{128 L^2
k^2 z^2} \left(1-f^h_c\int _0^{k z}e^{\frac{3}{2}c x^2
\left(-1+H_c(\frac{x}{k})\right)} x^3dx\right)\big[ 40k^2
z^2+\nonumber\\
&&64 c k^4 z^4-384 k^6 z^6+12\sqrt{2 \pi } e^{2 c k^2 z^2} k z
\big(-7+20 c k^2 z^2\big) Erf_c\big(\sqrt{2} k
z\big)\nonumber\\
&&-27 \pi  e^{4 c k^2 z^2} Erf_c\left(\sqrt{2} k
z\right){}^2\big]-\nonumber\\
&&\frac{3 f^h_c e^{\frac{5}{2}c k^2 z^2 \left(-1+H_c(z)\right)} k^3
z^3}{16 L^2 } \big[4 k z-16c k^3 z^3+3 \sqrt{2 \pi }e^{2 c k^2
z^2}Erf_c\big(\sqrt{2} k z\big)\big]\nonumber\\
\end{eqnarray}

We have used $V_E^{\pm}$ to represent the solution of the dilaton
potential for $c=\pm$. Where the function $\,
_2F_2\left(\{a_1,..,a_p\};\left\{b_1,..., b_q\right\};z\right)$ is a
generalized hypergeometric function which has series of $\sum
_{k=0}^{\infty } \left(a_1\right)_k\ldots
\left(a_p\right)_k/\left(b_1\right)_k\ldots \left(b_q\right)_k
\left.z^k\right/k!$ and $f_0$ is related to the integral constant
which can be expressed by the position of the horizon of the black
hole solution. The function $Erf_c[z]$ appeared is also a special
function class named by error function which is defined as integral
form $ Erf_+[z]=Erf[z]=\frac{2}{\sqrt{\pi }}\int _0^ze^{-t^2}dt$ and
$ Erf_-[z]=Erfi[z]=\frac{Erf[i z]}{i}$.

Conformal invariance in the UV can be obtained when $\phi\sim 0$ at
the UV boundary $z \rightarrow 0$. One can expand $\phi(z)$ at UV
boundary $z\sim 0$,
\begin{equation}
\label{asybehavior} \phi(z\rightarrow 0) \sim \frac{3c k^2
z^2}{2}+\frac{3 k^4 z^4}{10}+ \cdots .
\end{equation}
The behavior shown in Eq.(\ref{asybehavior}) is consistent with the
requirement of the asymptotic ${\rm AdS}_5$ near the ultraviolet
boundary.

Through the AdS/CFT dictionary, for any dilaton field $\Phi$, we
have
\begin{equation}
\label{confdim} \lim_{\Phi\rightarrow 0}V(\Phi) = -\frac{12}{L^2}
+\frac{1}{2L^2} \Delta(\Delta-4) \Phi^2 +O(\Phi^4).
\end{equation}
By using the below relationship,
\begin{eqnarray}
\partial^2_{\Phi} V(\Phi)=\frac{\partial r}{\partial \Phi}
\frac{\partial}{\partial r}\left(\frac{\partial r}{\partial \Phi}
\frac{\partial V(z) }{\partial r}\right)= M_{\Phi}^2 \Phi^2 + \cdots
,
\end{eqnarray}
one can easily get the conformal dimension of the dilaton field near
the UV boundary. In Eq. (\ref{confdim}), $\Delta$ is defined as
$\Delta( \Delta-4)= M_{\Phi}^2 L^2$, which is constrained by the
Breitenlohner-Freedman (BF) bound $2< \Delta< 4$. In our case,
$\Phi= \sqrt{\frac{8}{3}} \phi$,
\begin{equation}
M_{\Phi }^2=
 -\frac{4}{L^2},
\end{equation}
for $c=\pm$. Therefore, the conformal dimension of the dilaton field
is $\Delta=2$ in our holographic models with $c=\pm$, which
satisfies the BF bound but does not correspond to any local, gauge
invariant operator in 4D QCD. Although there have been some
discussions in recent years of the possible relevance of a dimension
two condensate in the form of a gluon mass term
\cite{Gubarev:2000eu}, it is not clear whether we can associate
$\phi$ with dimension-2 gluon condensate, because the $AdS/CFT$
correspondence requires that bulk fields should be dual to
gauge-invariant local operators.

\section{Equation of state for hQCD model with quadratic correction}
\label{section-eos}

A black-hole solution with a regular horizon is characterized by the
existence of a surface $z=z_h$, where $f(z_h)=0$. The Euclidean
version of the solution is defined only for $0 < z < z_h$, in order
to avoid the conical singularity, the periodicity of the Euclidean
time can be fixed by
\begin{equation}
\tau \rightarrow \tau+\frac{4\pi}{|f'(z_h)|}.
\end{equation}
This determines the temperature of the solution as
\begin{equation}
T=\frac{|f'(z_h)|}{4\pi}.
\end{equation}

From Eq. (\ref{solu-f}), one can easily read out the relation
between the temperature and position of the black hole horizon.
\begin{equation} \label{temp}
T = \frac{k^4 z_h^3 \exp \left(\frac{3}{2} \left(c k^2 z_h^2 \,
H_c(z_h)- ck^2 z_h^2\right)\right)}{4 \pi \int_0^{k z_h}
e^{\frac{3}{2} \left(-ck^2 x^2+ck^2 x^2 H_c(x)+\text{Log}\left[k^2
x^2\right]\right)}  \, dx}
\end{equation}

\subsection{The would-be critical temperature}
\label{section-Tc}

\begin{figure}[h]
\begin{center}
\epsfxsize=6.5 cm \epsfysize=6.5 cm \epsfbox{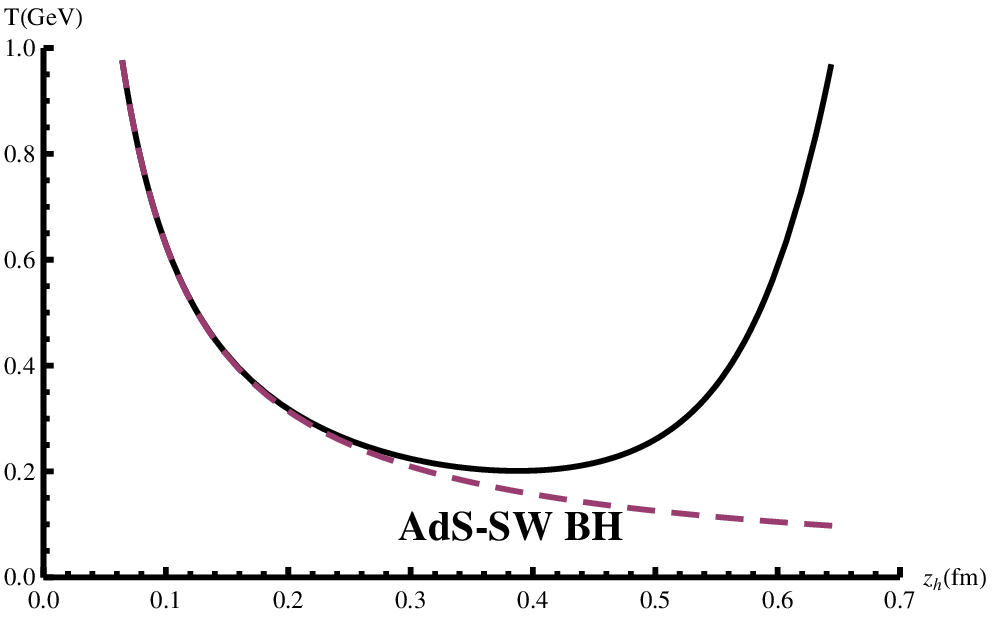}
\hspace*{0.1cm} \epsfxsize=6.5 cm \epsfysize=6.5 cm
\epsfbox{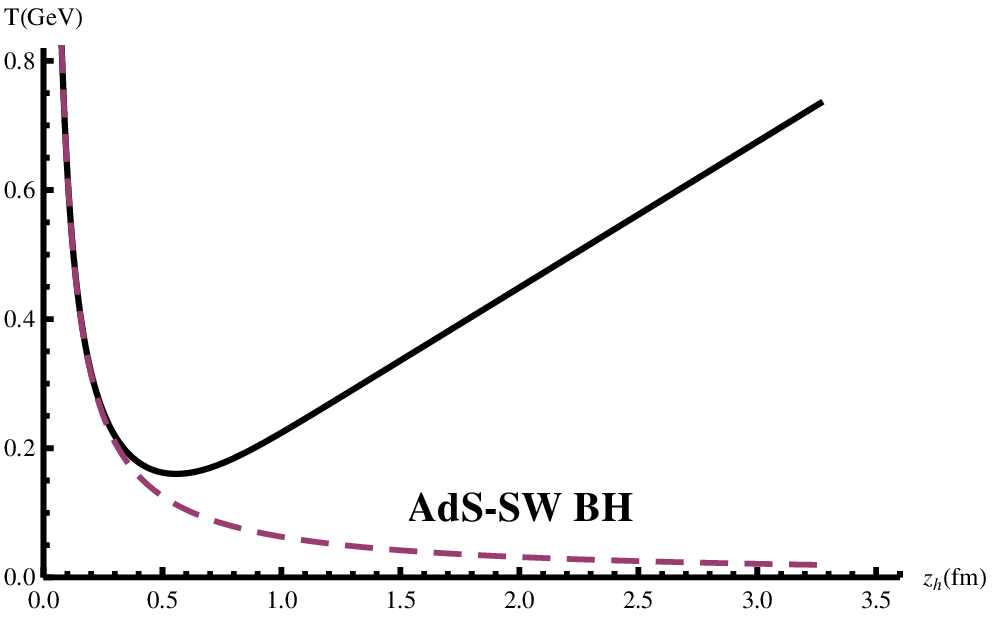} \vskip -0.05cm \hskip 0.15 cm
\textbf{( a ) } \hskip 6.5 cm \textbf{( b )} \\
\end{center}
\caption{The temperature $T$ as a function of the black-hole horizon
$z_h$ with $k=0.43 GeV$ for $c=+$ in (a) and $c=-$ in (b),
respectively. The dashed lines are for pure ${\rm AdS}_5$ Schwarz
black-hole. }
 \label{Tem-rh}
\end{figure}

According to Eq.(\ref{temp}), we cannot get the analytical
expression to describe the relation between the temperature and
horizon. The numerical results for the temperature as a function of
the horizon $z_h$ for $c=+$ and $c=-$ are shown in Fig.\ref{Tem-rh}
(a) and (b), respectively. Here we have taken $k=0.43 {\rm GeV}$.
The dashed lines are results from the pure ${\rm AdS}_5$ Schwarz
black-hole in Eq.(\ref{pureAdSBH}).

From Fig. \ref{Tem-rh}, it is noticed that for pure ${\rm AdS}_5$
Schwarz black-hole, the temperature monotonically decreases with the
increase of the horizon \cite{Andreev-T1,Andreev-T2, Andreev-T3}. If
one solves the dual black-hole background self-consistently, one can
find that for both hQCD models with positive and negative quadratic
correction, there is a minimal temperature $T_{min}$ at certain
black-hole horizon $z_h^0$. This is similar to the case for the
confining theory (at zero temperature) discussed in Ref.
\cite{Gursoy-T1}. For $T<T_{min}$, there are no black-hole
solutions. For $T>T_{min}$, there are two branches of black-hole
solutions. When $z_h<z_h^0$, the temperature increases with the
decreasing of $z_h$, which means that the temperature increases when
the horizon moves close to UV, this phase is thermodynamically
stable. When $z_h>z_h^0$, the temperature increases with the
increase of $z_h$, which means that the temperature becomes higher
and higher when the horizon moves to IR. This indicates that the
solution for the branch $z_h>z_h^0$ is unstable and thus not
physical.

In order to determine the critical temperature, we have to compare
the free energy difference between the stable black hole solution
and the thermal gas. The thermal gas is solved by setting $f_0(z)=1$
\cite{Gursoy-T1,Herzog}. From the AdS/CFT conjecture, the gravity
side is weakly coupled, so we can use the semi-classical
approximation and just deal with the on-shell action. Following Ref.
\cite{Herzog} and Appendix C in Ref. \cite{Gursoy-T1}, the
regularized total on-shell action in Euclidean space can be
decomposed as three parts:
\begin{equation}
I^R=I^R_{E}+I^R_{GH}+I^R_{count}=I^{\epsilon} + I^{IR}+ I^{count},
\end{equation}
with $I^R_E$ the on-shell Einstein action, $I^R_{GH}$ the
Gibbons-Hawking term and $I^R_{count}$ the counter term from the
renormalization scheme. $I^{\epsilon}$ and $I^{IR}$ are the
contributions at UV cut-off $z=\epsilon $ and IR cut-off $z=z_{IR}$,
respectively. Correspondingly, the free energy density is defined as
\begin{equation}
{\cal F}=\frac{T}{V_3}I^R={\cal F}^{\epsilon}+{\cal F}^{IR}+{\cal
F}^{count},
\end{equation}
with $V_3$ the volume of the three-space.

To compare the free-energy density of the black-hole case and the
thermal gas case, one has to impose the following conditions
\cite{H-H}:
\begin{eqnarray}
\phi_0(z)\mid_{\epsilon}=&& \phi(z)\mid_{\epsilon},\\
\tilde{\beta} b_0(z)\mid_{\epsilon}=&&\beta b(z)\sqrt{f(z)}\mid_{\epsilon},\\
\tilde{V}_3 b_0^3(z)\mid_{\epsilon}=&&V_3 b^3(z)\mid_{\epsilon}.
\end{eqnarray}
Where $\phi_0, b_0,\tilde{\beta},\tilde{V}_3$ are for thermal gas
with $b_0=\frac{L}{z} e^{A_{E0}(z)}$ and
$A_{E0}=A_{s0}-\frac{2}{3}\phi_0$, and $\phi, b,\beta,V_3$ are for
the black-hole with $b(z)=\frac{L}{z} e^{A_E(z)}$.

The free energy density for the black-hole has the form of
\begin{eqnarray}
{\cal F}_{BH}&=& {\cal F}_{BH}^{\epsilon}+{\cal F}_{BH}^{count} \\
{\cal F}_{BH}^{\epsilon}&= & 2M^3 \left(3 b(\epsilon )^2 f(\epsilon
) b'(\epsilon )+\frac{1}{2} b(\epsilon )^3 f'(\epsilon )\right),
\end{eqnarray}
where we have defined $M^3=1/(16 \pi G_5)$. It is noticed that for
black-hole case, $z_{IR}$ is normally set at the horizon $z_h$,
therefore $I^{IR}$ vanishes due to $f(z_h)=0$. At the limit
$\epsilon\rightarrow 0$, we have
\begin{equation}
{\cal F}_{BH}^{\epsilon}=-\frac{6 \left(L^3 M^3\right)}{\epsilon
^4}-\left(\frac{6}{5} k^4 L^3 M^3-\frac{1}{2} k^4 L^3 M^3
f_c^h\right)+o(\epsilon^2). \label{f-BH-UV}
\end{equation}
For detailed derivation of Eq.(\ref{f-BH-UV}), please refer to
Appendix \ref{appendix-epsilon}.

The free energy density for the thermal gas takes the form of
\begin{eqnarray}
{\cal F}_{TG}&=& {\cal F}_{TG}^{\epsilon}+{\cal F}_{TG}^{IR} + {\cal
F}_{TG}^{count}, \\
{\cal F}_{TG}^{\epsilon}&=& 2 M^3
\frac{b(\epsilon )^4 \sqrt{f(\epsilon)}} {b_0(\epsilon )^4
\sqrt{f_0(\epsilon )}} \left(3 b_0(\epsilon )^2 f_0(\epsilon )
b_0'(\epsilon )+\frac{1}{2}
b_0(\epsilon )^3 f_0'(\epsilon )\right), \\
{\cal F}_{TG}^{IR}&=&
2M^3f_0(z_{IR})b_0^2(z_{IR})b_0^\prime(z_{IR}).
\end{eqnarray}
At the limit $\epsilon\rightarrow 0$, from the results derived in
Appendix \ref{appendix-epsilon}, the free energy density of the
thermal gas at UV cut-off takes the form of
\begin{equation}
{\cal F}_{TG}^{\epsilon}=-\frac{6 L^3
M^3}{\epsilon^4}-\left(\frac{6}{5} k^4 L^3 M^3-\frac{3}{4} k^4 L^3
M^3 f_c^h\right)+o(\epsilon^2).
\end{equation}
Where we have used $f_0(z)=1$ in the derivation.

The divergent behaviors of the thermal gas and black-hole are the
same, i.e, the counter terms ${\cal F}_{BH}^{count}$ and ${\cal
F}_{TG}^{count}$ would have the same contribution in the two phases
thus ${\cal F}_{BH}^{count}-{\cal F}_{TG}^{count}$ would disappear
in the free energy difference. Therefore, the final expression of
the free energy difference has the form of
\begin{eqnarray}
\Delta {\cal F}={\cal F}_{BH}-{\cal F}_{TG}=&&-\frac{1}{4} k^4 M^3
L^3 f_c^h-2M^3 b_0^2(z_{IR})b_0^\prime(z_{IR}).
\label{d-free-energy}
\end{eqnarray}
It is worthy of mentioning that $f_c^h>0$, therefore the
contribution from the first term is always negative. Furthermore,
from Eq.(\ref{dF-s}), we can observe that the first term
contribution is exactly $-\frac{1}{4}Ts$ with $s$ the entropy
density, and the free energy difference can be written as:
\begin{equation}
\Delta {\cal F}=-\frac{1}{4}Ts-2M^3b_0^2(z_{IR})b_0^\prime(z_{IR}).
\label{dF-sT-zIR}
\end{equation}
Because $b_0(z)$ is monotonically decreasing with $z$ so
$b_0^\prime(z)<0$, the second term contribution to the free energy
difference is always positive. The expression of
Eq.(\ref{dF-sT-zIR}) is similar to Eq.(3.25) in
Ref.\cite{Gursoy-T1}. We can observe that the second term
contribution in Eq.(\ref{dF-sT-zIR}) plays the same role of the
gluon-condensate in Ref.\cite{Gursoy-T1}.

\begin{figure}[h]
\begin{center}
\epsfxsize=6.5 cm \epsfysize=6.5 cm \epsfbox{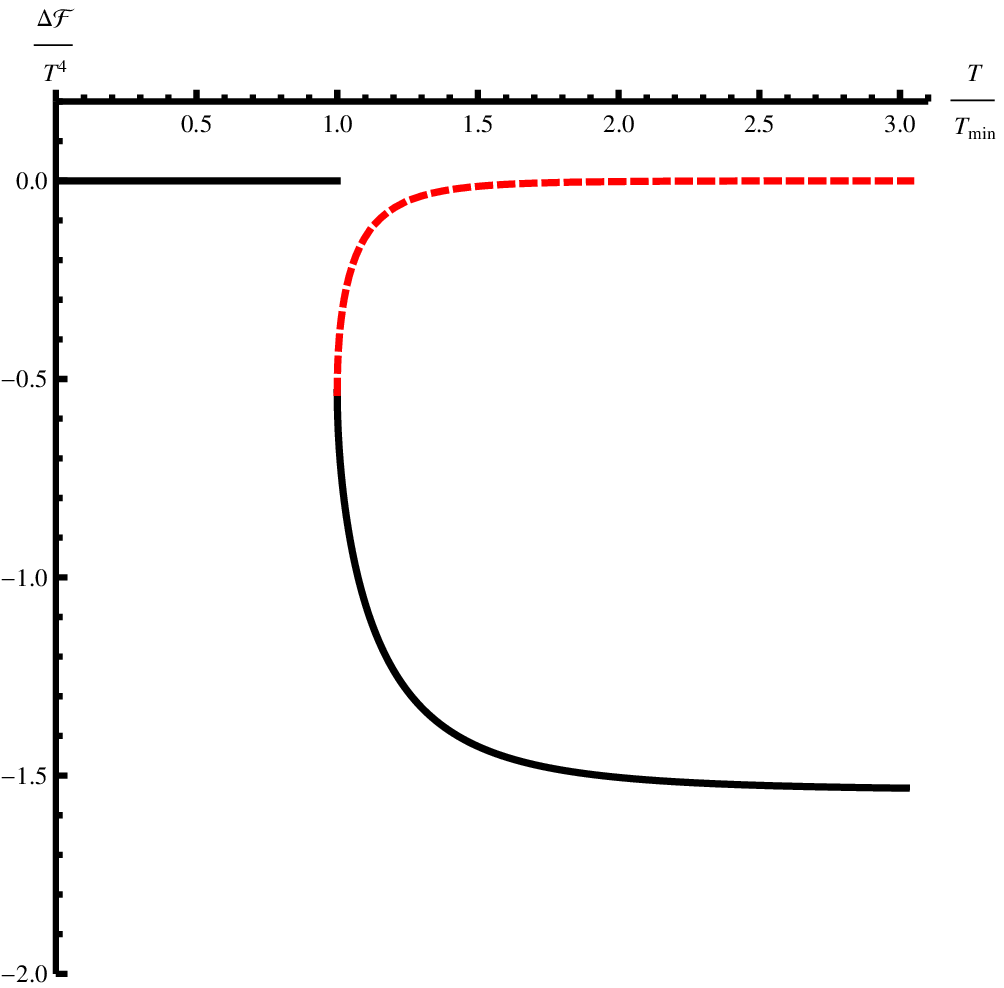} \hspace*{0.1cm}
\epsfxsize=6.5 cm \epsfysize=6.5 cm \epsfbox{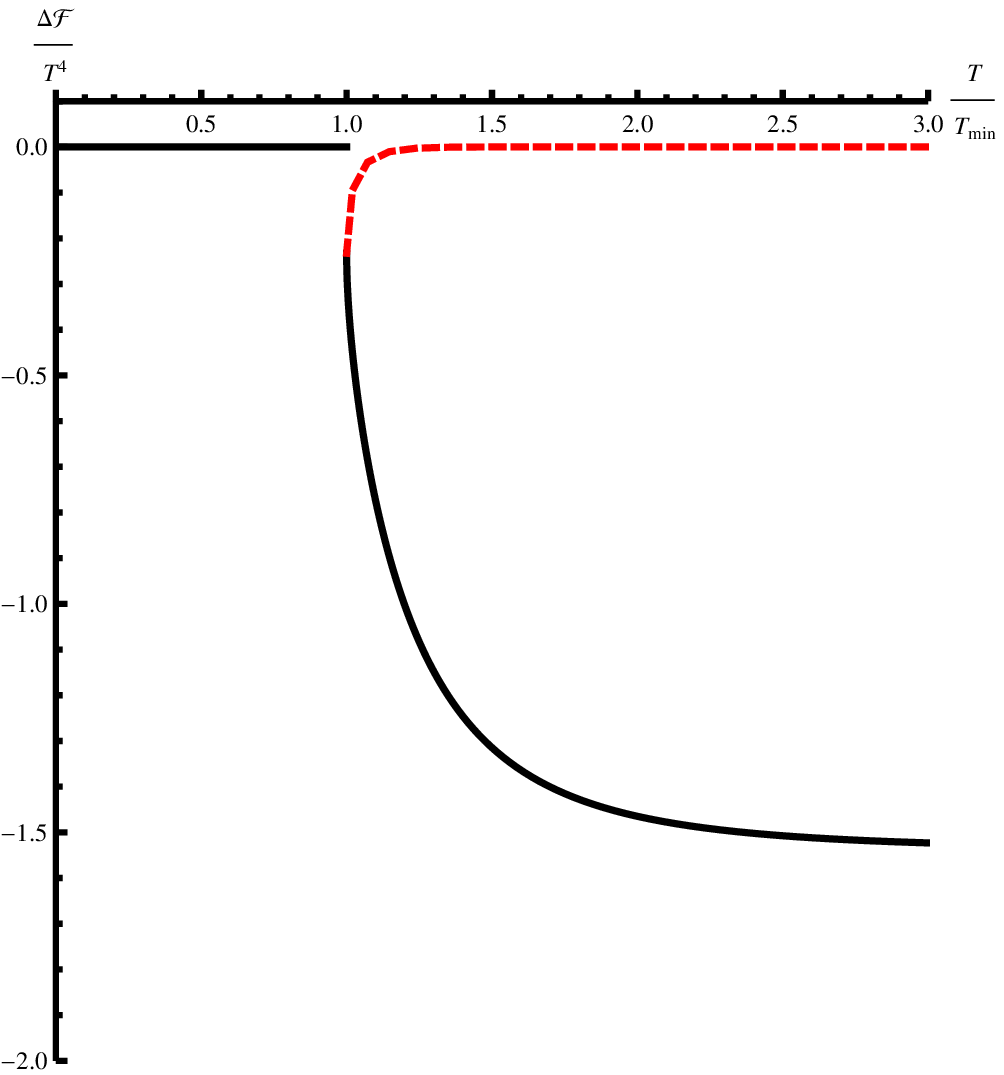} \vskip -0.05cm
\hskip 0.15 cm
\textbf{( a ) } \hskip 6.5 cm \textbf{( b )} \\
\end{center}
\caption[]{The free energy difference of
Eq.(\ref{d-free-energy-IRinfty}) scaled by $T^4$ as a function of
$T/T_{min}$ with $k=0.43 GeV$ and $G_5/L^3=1.26$ for $c=+$ in (a)
and $c=-$ in (b), respectively. The dashed lines are for unstable
black-hole solution. } \label{Tem-dF}
\end{figure}

The second term contribution in Eq.(\ref{dF-sT-zIR}) is $z_{IR}$
dependent. If we follow Ref.\cite{Gursoy-T1} and choose $z_{IR}$ at
the good singularity point, i.e, $z_{IR}\rightarrow \infty $, where
$b_0(z_{IR})\rightarrow 0$, then the second term contribution in
Eq.(\ref{d-free-energy}), i.e, the $z_{IR}$ related term vanishes
and we have
\begin{eqnarray}
\Delta {\cal F}={\cal F}_{BH}-{\cal F}_{TG}=&&-\frac{1}{4} k^4 M^3
L^3 f_c^h = -\frac{1}{4}Ts < 0. \label{d-free-energy-IRinfty}
\end{eqnarray}
This indicates that the black-hole phase is more stable than the
thermal gas. The free energy difference for the case of $z_{IR}
\rightarrow \infty $ as a function of $T/T_{min}$ is shown in Fig.
\ref{Tem-dF}. The solid lines indicate the stable black-hole
solution, and the dashed lines are for the unstable black-hole
phase. It is seen that $\Delta {\cal F}$ jumps to a negative value
at $T_{min}$, then decreases with the increase of the temperature.
In this case, we observe a zeroth order phase transition at
$T_{min}$ due to the free energy discontinuity. By choosing the
parameters $k=0.43 {\rm GeV}$ and $G_5/L^3=1.26$, the value of
$-\Delta {\cal F}/T^4$ approaches $1.5$ at high temperature, which
is similar to the lattice data of the pressure density for pure
SU(3) gauge theory \cite{LAT-EOS-G}.

In principle, the critical temperature $T_c$ from
Eq.(\ref{dF-sT-zIR}) should be determined by finite value of
$z_{IR}$, which sets the basic scale of the theory and should be
fixed by experimental or lattice data. For example, for
$z_{IR}=2.34{\rm GeV}^{-1}$, the free-energy difference will cross
the zero-energy axis at $T_{min}$ so that $T_c=T_{min}$ and the
phase transition will be of first order. If we choose other smaller
value of $z_{IR}$, e.g, $z_{IR}= 2 {\rm GeV}^{-1}$, the free energy
density difference will cross the zero-energy axis at $T_c
>T_{min}$ and the phase transition will be also of 1st order.

For our numerical calculations, we will firstly choose
$T_{min}(z_h^0) $ as the "would-be" critical temperature, which
means that we choose a finite value of $z_{IR}=2.34{\rm GeV}^{-1}$
and make $\delta{\cal F}=0$ at $T_{min}$. The exact thermal gas
solution in the region $T<T_{min}$ is unknown for us, therefore, we
only focus on the temperature region $T>T_{min}$ for the deconfined
quark gluon plasma phase. In the following we will compare our
results on the equation of state as a function of $T/T_{min}$ with
the lattice data as a function of $T/T_c$.

The value of $T_{min}$ is only dependent on the model parameter $k$
in Eq.(\ref{ourmodel}). With given $k=0.43 {\rm GeV}$, we can read
the values of the critical temperature $T_{min}=201 {\rm MeV}$ and
$T_{min}=160 {\rm MeV}$ for $c=+$ and $c=-$, respectively. Both
values are in agreement with lattice QCD result on the critical
temperature \cite{LAT-EOS-G, LAT-EOS-Nf2}.

\subsection{The entropy density}

\begin{figure}[h]
\begin{center}
\epsfxsize=6.5 cm \epsfysize=6.5 cm \epsfbox{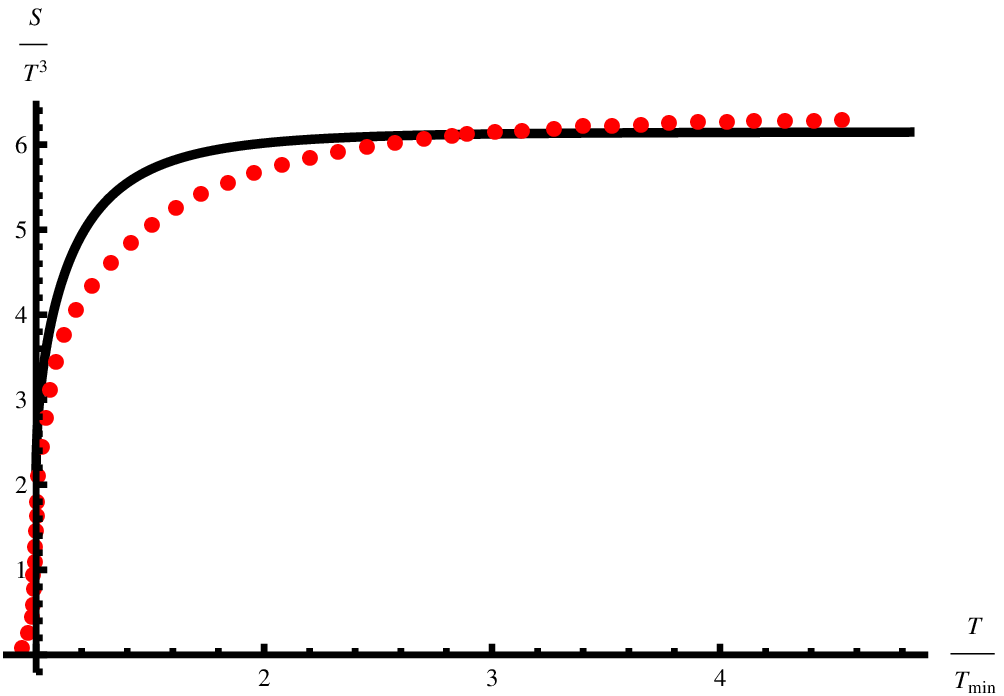}
\hspace*{0.1cm} \epsfxsize=6.5 cm \epsfysize=6.5 cm
\epsfbox{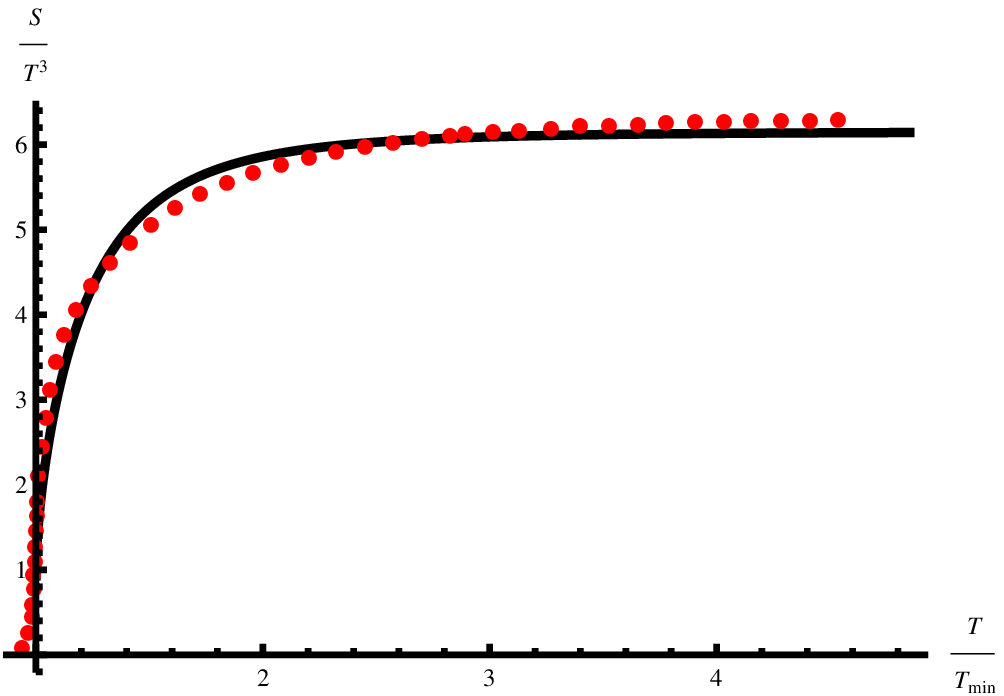} \vskip -0.05cm \hskip 0.15 cm
\textbf{( a ) } \hskip 6.5 cm \textbf{( b )} \\
\end{center}
\caption{The scaled entropy density $s/T^3$ as a function of scaled
temperature $T/T_{min}$ with $k=0.43 {\rm GeV}$ and $G_5/L^3=1.26$
for $c=+$ in (a) and $c=-$ in (b), respectively. The dotted lines in
(a) and (b) are lattice results as a function of $T/T_c$ from
\cite{LAT-EOS-G}.} \label{EntyDensity}
\end{figure}

Following the standard Bekenstein-Hawking formula \cite{entropy-BK},
from the geometry given in Eq.(\ref{metric-Einsteinframe}), one can
easily read the black-hole entropy density $s$, which is defined by
the area $A_{area}$ of the horizon:
\begin{equation}
\label{entrpy} s=\frac{A_{area}}{4 G_5 V_3}|_{z_h}=
\frac{L^3}{4G_5}\left(\frac{e^{A_s-\frac{2}{3}\phi}}{z}\right)^3|_{z_h}.
\end{equation}
Where $G_5 $ is the Newton constant in 5D curved space and $V_3$ is
the volume of the spatial directions. It is noticed that the entropy
density is closely related to the metric in the Einstein frame.

With fixed $k=0.43 {\rm GeV}, G_5/L^3=1.26$, the scaled entropy
density $s/T^3$ as a function of scaled temperature $T/T_{min}$ is
shown in Fig.\ref{EntyDensity}(a) for $c=+$ ($T_{min}=201 {\rm
MeV}$) and Fig.\ref{EntyDensity}(b) for $c=-$ ($T_{min}=160 {\rm
MeV}$), respectively. The dots in Fig.\ref{EntyDensity} are the
lattice result taken from \cite{LAT-EOS-G}. It is can be seen that
the entropy density in both positive and negative quadratic
correction hQCD models agrees well with the lattice result for pure
$SU(3)$ gauge theory.

\subsection{The pressure density, energy density and trace anomaly}

The pressure density $p(T)$ can be calculated from the entropy
density $s(T)$ by solving the equation
\begin{equation}
\frac{dp(T)}{dT}= s(T).
\end{equation}

\begin{figure}[h]
\begin{center}
\epsfxsize=6.5 cm \epsfysize=6.5 cm \epsfbox{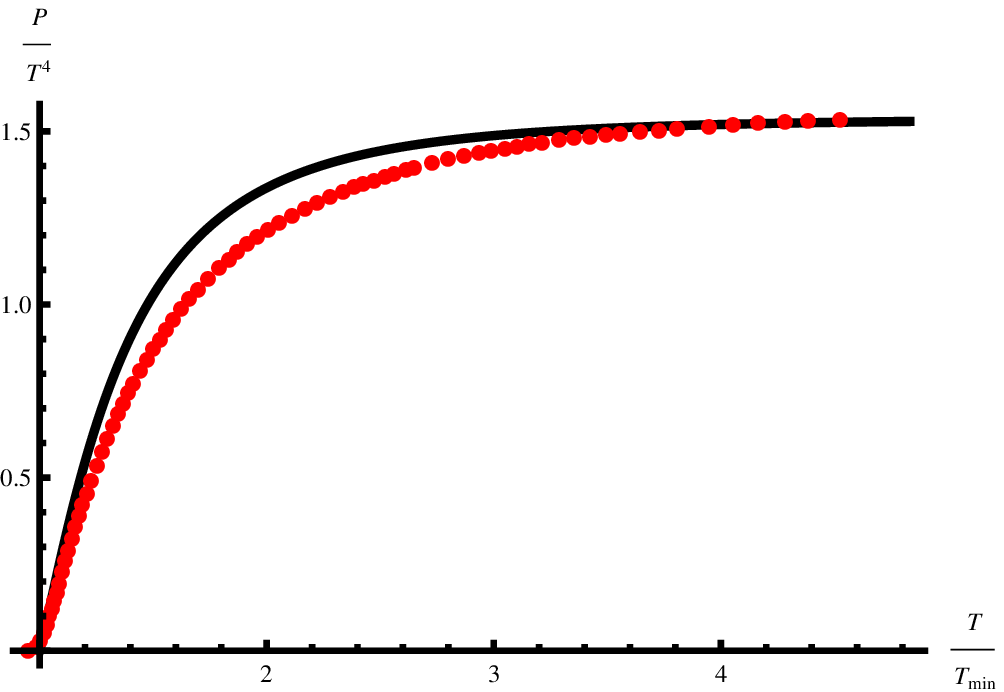}
\hspace*{0.1cm} \epsfxsize=6.5 cm \epsfysize=6.5 cm
\epsfbox{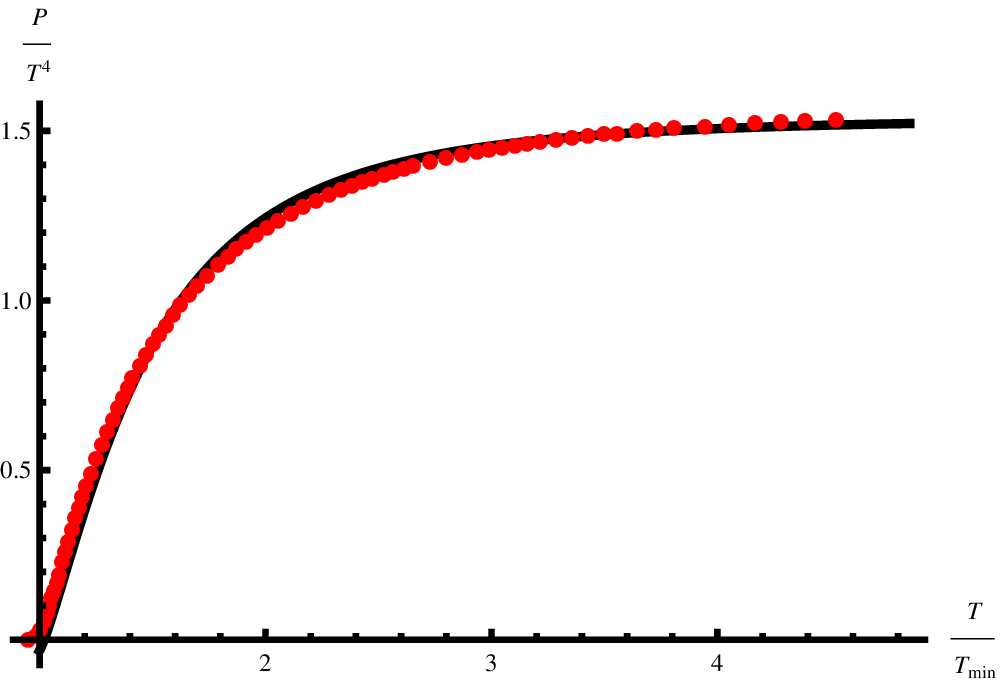} \vskip -0.05cm \hskip 0.15 cm
\textbf{( a ) } \hskip 6.5 cm \textbf{( b )} \\
\end{center}
\caption{The scaled pressure density $p/T^4$ as a function of scaled
temperature $T/T_{min}$ with $k=0.43 GeV$ and $G_5/L^3=1.26$ for
$c=+$ in (a) and $c=-$ in (b), respectively. The dots are the
lattice data as a function of $T/T_c$ from \cite{LAT-EOS-G}.}
\label{Pressure}
\end{figure}

After integrating the Bekenstein-Hawking entropy density, the
pressure density of the system can be obtained up to a integral
constant $p_0$. In our numerical calculation, we have set $p_0=0$ to
ensure $p(T_{min})=0$. In some sense, this procedure is equivalent
to choose a finite value of $z_{IR}$ in Eq.(\ref{dF-sT-zIR}) to
ensure $\Delta{\cal F}=0$ at $T_{min}$.

The numerical result of the pressure density as a function of
temperature is shown in Fig.\ref{Pressure}. With fixed $k=0.43 {\rm
GeV}, G_5/L^3=1.26$, the scaled pressure density $p/T^4$ as a
function of scaled temperature $T/T_{min}$ is shown in
Fig.\ref{Pressure}(a) for $c=+$ ($T_{min}=201 {\rm MeV}$) and
Fig.\ref{Pressure}(b) for $c=-$ ($T_{min}=160 {\rm MeV}$),
respectively. The dots in Fig.\ref{Pressure} are the lattice result
of pressure density for pure SU(3) gauge theory \cite{LAT-EOS-G}. It
can be seen that the pressure density in both hQCD models with
$c=\pm$ agree well with the lattice result.

\begin{figure}[h]
\begin{center}
\epsfxsize=6.5 cm \epsfysize=6.5 cm \epsfbox{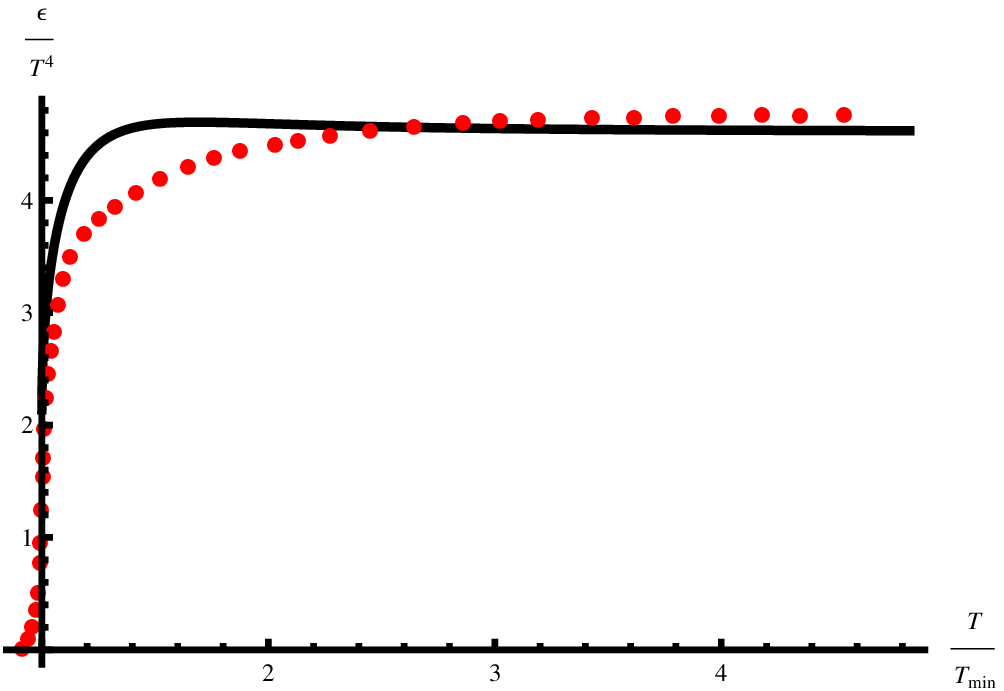}
\hspace*{0.1cm} \epsfxsize=6.5 cm \epsfysize=6.5 cm
\epsfbox{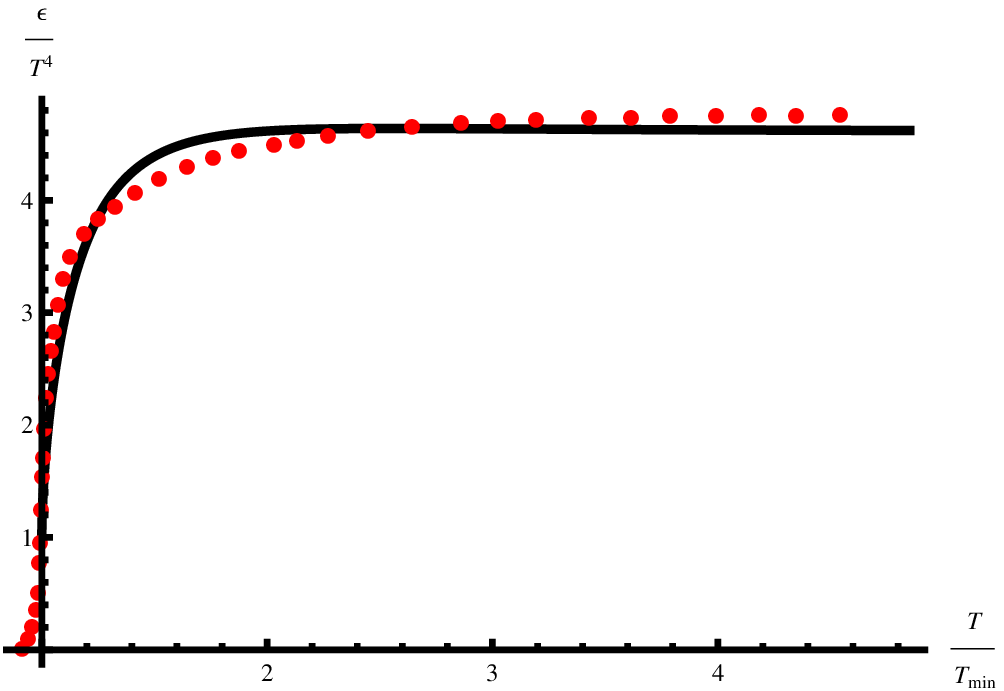} \vskip -0.05cm \hskip 0.15 cm
\textbf{( a ) } \hskip 6.5 cm \textbf{( b )} \\
\end{center}
\caption{The scaled energy density $\epsilon/T^4$ as a function of
scaled temperature $T/T_{min}$ with $k=0.43 GeV$ and $G_5/L^3=1.26$
for $c=+$ in (a) and $c=-$in (b), respectively. The dots are lattice
data as a function of $T/T_c$ from \cite{LAT-EOS-G}. }
\label{EnergyDen}
\end{figure}

Once we get the entropy density and pressure, we can get the energy
density $\epsilon$ immediately. The energy density is defined by
\begin{equation}
\epsilon=-p+ s T.
\end{equation}
The numerical result of energy density as a function of the
temperature is shown in Fig.\ref{EnergyDen}. The lattice data for
pure $SU(3)$ gauge theory is shown with dots. It is found that the
energy density from the two hQCD models with $c=\pm$ agrees with the
lattice data.

\begin{figure}[h]
\begin{center}
\epsfxsize=6.5cm \epsfysize=6.5cm \epsfbox{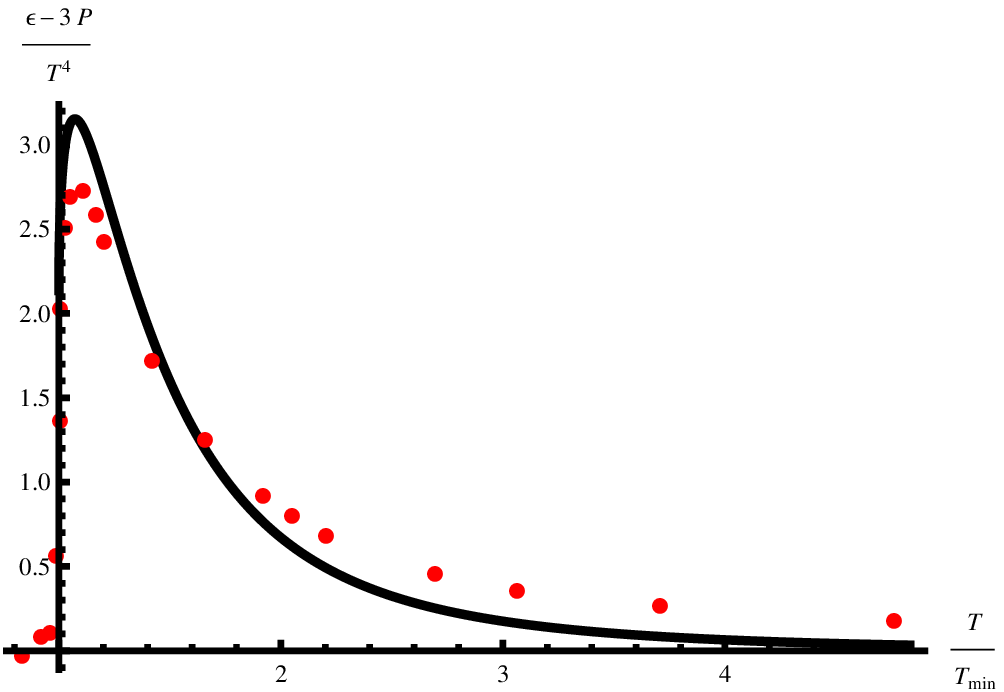}
\hspace*{0.1cm} \epsfxsize=6.5 cm \epsfysize=6.5 cm
\epsfbox{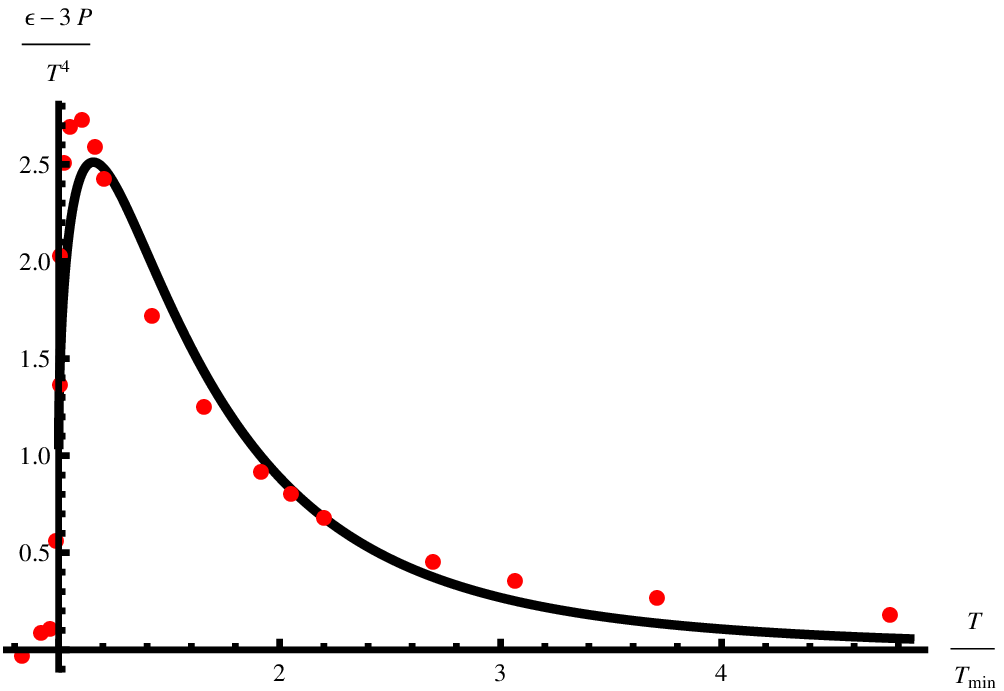} \vskip -0.05cm \hskip 0.15 cm
\textbf{( a ) } \hskip 6.5 cm \textbf{( b )} \\
\end{center}
\caption{The trace anomaly $\epsilon-3p$ as a function of scaled
temperature $T/T_{min}$ with $k=0.43 GeV$ and $G_5/L^3=1.26$ for
$c=+$ in (a) and $c=-$ in (b), respectively. The dots are the
lattice data as a function of $T/T_c$ from \cite{LAT-EOS-G}. }
\label{TraceA}
\end{figure}

We also show the trace anomaly $\epsilon-3p$ in Fig.\ref{TraceA}.
The trace anomaly shows a peak around $T/T_{min}=1.1$ for the case
of $c=+$ and $T/T_{min}=1.2$ for the case of $c=-$, and the height
is around $3$ for $c=+$ and $2.5$ for $c=-$. Both results are in
agreement with lattice result. At very high temperature, the trace
anomaly goes to zero, which indicates the system is asymptotically
conformal at high temperature.

\subsection{The sound velocity and specific heat}

\begin{figure}[h]
\begin{center}
\epsfxsize=6.5 cm \epsfysize=6.5 cm \epsfbox{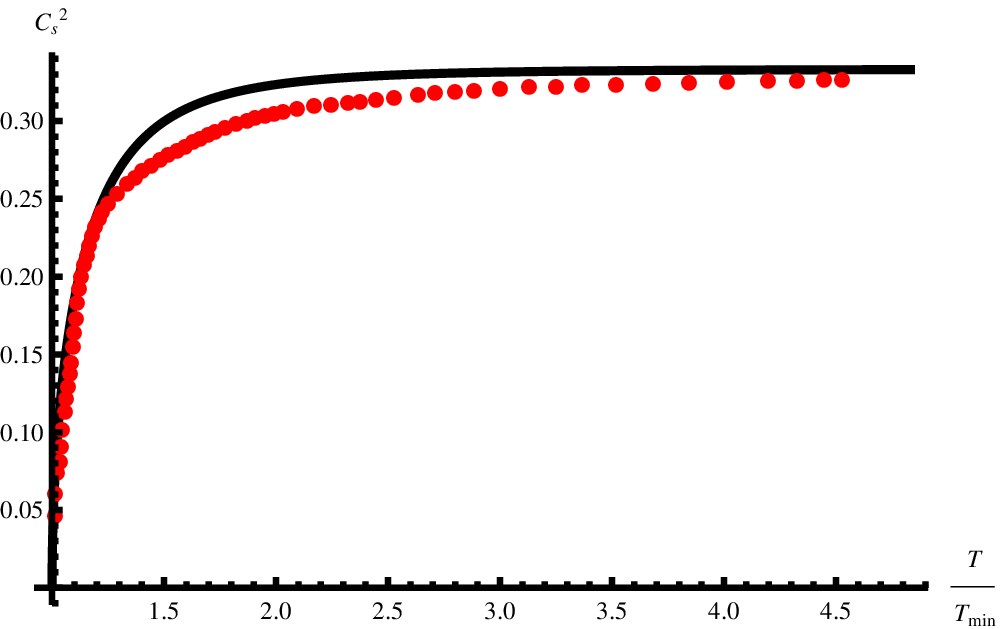}
\hspace*{0.1cm} \epsfxsize=6.5 cm \epsfysize=6.5 cm
\epsfbox{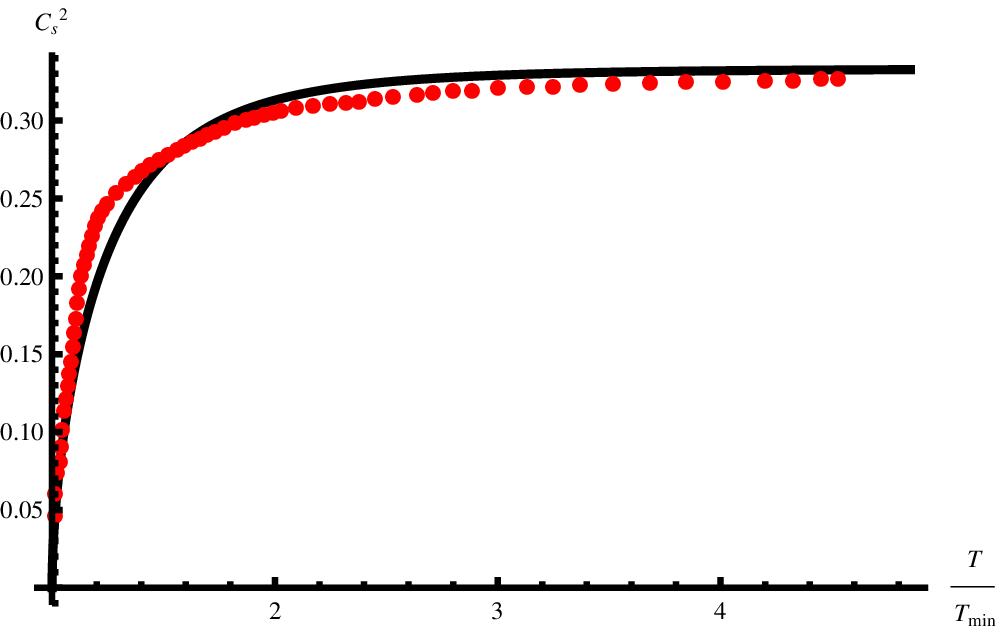} \vskip -0.05cm \hskip 0.15 cm
\textbf{( a ) } \hskip 6.5 cm \textbf{( b )} \\
\end{center}
\caption{The square of the sound velocity $c_s^2$ as a function of
scaled temperature $T/T_{min}$ with $k=0.43 GeV$ and $G_5/L^3=1.26$
for $c=+$ ($T_{min}=201 {\rm MeV}$) and $c=-$ ($T_{min}=160 {\rm
MeV}$), respectively. The dotted line is the lattice data as a
function of $T/T_c$ from \cite{LAT-EOS-G}.}
 \label{CS}
\end{figure}

\begin{figure}[h]
\begin{center}
\epsfxsize=6.5 cm \epsfysize=6.5 cm \epsfbox{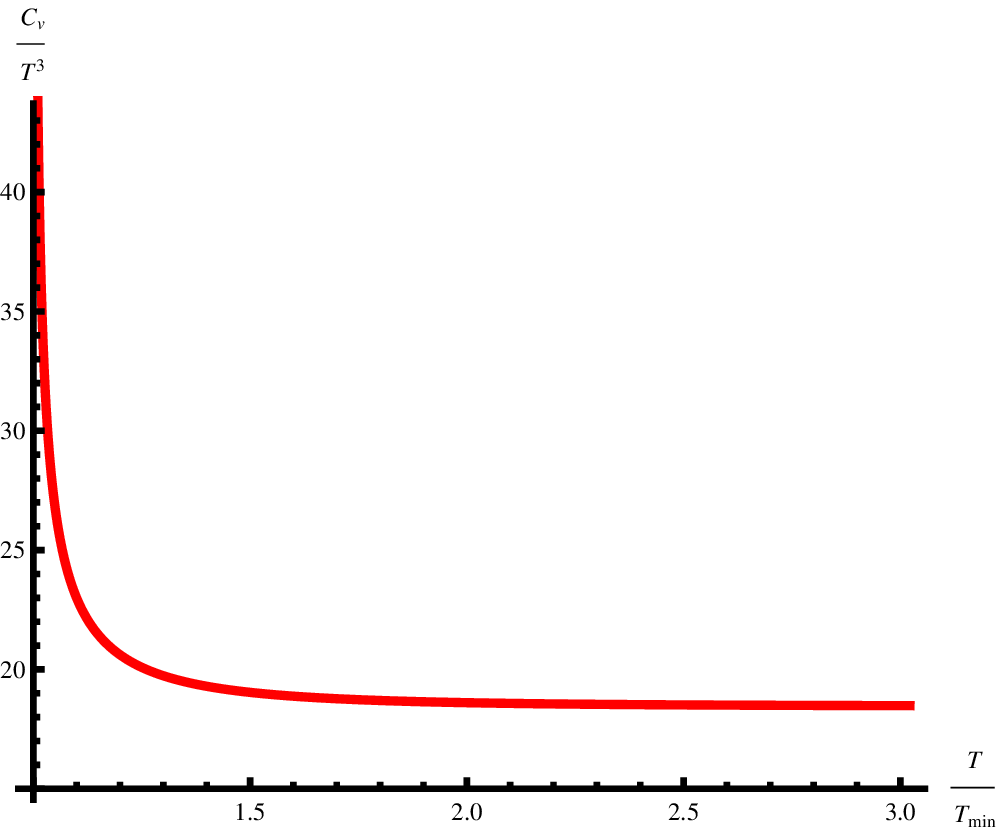}
\hspace*{0.1cm} \epsfxsize=6.5 cm \epsfysize=6.5 cm
\epsfbox{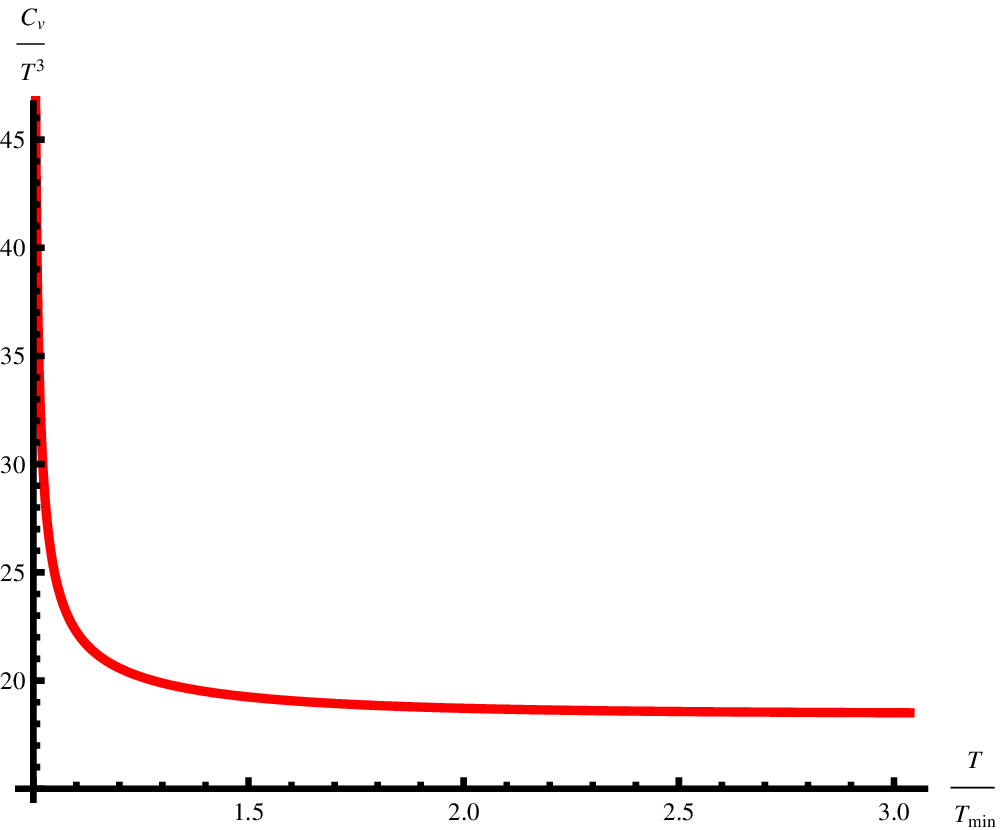} \vskip -0.05cm \hskip 0.15 cm
\textbf{( a ) } \hskip 6.5 cm \textbf{( b )} \\
\end{center}
\caption{The scaled specific heat $C_v/T^3$ as a function of scaled
temperature $T/T_{min}$ with $k=0.43 GeV$ and $G_5/L^3=1.26$ for
$c=+$ ($T_{min}=201 {\rm MeV}$) and $c=-$ ($T_{min}=160 {\rm MeV}$),
respectively. }
 \label{CV}
\end{figure}

The sound velocity $c_s^2$ can be obtained from the temperature and
entropy:
\begin{equation}\label{sound}
c_s^2=\frac{d \log T}{d \log s}=\frac{s}{T ds/dT},
\end{equation}
which can directly measure the conformality of the system. For
conformal system, $c_s^2=1/3$, for non-conformal system, $c_s^2$
will deviate from $1/3$. From Eq.(\ref{sound}), we can see that the
speed of the sound is independent of the normalization of the 5D
Newton constant $G_5$ and the space volume $V_3$.

With known entropy density and the speed of sound, the specific heat
$C_v$ can be derived straightforwardly and takes the following
expression:
\begin{equation}\label{specificheat}
C_v=\frac{d \epsilon}{d T}=T \frac{d s}{d T}=s/c_s^2.
\end{equation}
At phase transition, the specific heat shows a clear $\lambda$-type
anomaly or divergent behavior \cite{landau-book}, which is
consistent with the jump or fast change of entropy density at $T_c$.
Therefore, it can be used to determine the phase transition point.
At very high temperature, the specific heat for the free gas of pure
gauge $SU(N_c)$ theory has the relation \cite{CV-lattice}
\begin{equation}\label{cv-freegas}
C_v/T^3 = 4\epsilon/T^4 \rightarrow 4 (N_c^2-1)\pi^2/15,
\end{equation}
and at large $N_c$ limit \cite{Gursoy-T1}, it takes the limit of
\begin{equation}\label{cv-freegas-largeNc}
C_v/T^3 \rightarrow 4 N_c^2\pi^2/15.
\end{equation}

The numerical result of the square of the sound velocity is shown in
Fig.\ref{CS}. At $T_{min}$, the sound velocity square is around
$0.05$ for $c=+$ and $0.02$ for $c=-$, both are in agreement with
lattice data $0.05$. At high temperature, the sound velocity square
goes to $1/3$, which means that the system is asymptotically
conformal. The numerical result of the specific heat is shown in
Fig.\ref{CV}. It can be clearly seen that the specific heat $C_v$
diverges at $T_{min}$.  At $T\rightarrow \infty$, we can see that
the scaled specific heat $C_v/T^3$ approaches the free gas limit of
pure gluon system $4 (N_c^2-1)\pi^2/15$ with $N_c=3$.

\subsection{Discussion}

In this section, we have investigated the equation of state for two
hQCD models with positive/negative quadratic correction in the
deformed warp factor of ${\rm AdS_5}$. It has been found that for
both models, if we choose $T_c=T_{min}$, the properties of entropy
density, pressure density, energy density and sound velocity are all
in agreement with lattice results for pure SU(3) gauge theory. As we
have mentioned, $T_c$ can be larger than $T_{min}$. By comparing our
numerical results with lattice data, it is observed that to fit the
lattice data well, $T_c$ should be located in a neighborhood of
$T_{min}$ from above, which means $z_{IR}$ should be in a
neighborhood of $z_{IR}=2.34 {\rm GeV}^{-1}$ from below. If $T_c$
lies in this small region around $T_{min}$, the numerical results on
equation of state can fit lattice data almost equally well as that
at $T_c=T_{min}$. It is worthy of mentioning that in our work
choosing $T_c=T_{min}$ can fit lattice data better. Therefore, we
identify $T_{min}$ as $T_c$ in this work, and we believe that the
qualitative results are not dependent on the exact location of
$T_c$.

\section{Heavy quark potential, Polyakov loop, spatial string tension at finite temperature}
\label{section-loop}

QCD vacuum is characterized by spontaneous chiral symmetry breaking
and color confinement. The dynamical chiral symmetry breaking is due
to a non-vanishing quark anti-quark condensate, $\langle{\bar q}q
\rangle \simeq (250 {\rm MeV})^3$ in the vacuum, which induces the
presence of the light Nambu-Goldstone particles, the pions and kaons
in the hadron spectrum. The confinement represents that only
colorless states are observed in the spectrum, which is commonly
described by the linearly rising potential between two heavy quarks
at large distances, $V_{\bar{Q}Q}(r)=\sigma r$, where $r$ is the
distance between quark anti-quark, and $\sigma \simeq (425 {\rm
MeV})^2$ is the string tension of the flux tube.

It is expected that chiral symmetry can be restored and color
degrees of freedom can be freed at high temperature and/or density.
The chiral restoration and deconfinement phase transitions are
characterized by the breaking and restoration of chiral and center
symmetry, which are only well defined in two extreme quark mass
limits, respectively. In the chiral limit when the current quark
mass is zero $m=0$, the chiral condensate $\langle{\bar q}q \rangle$
is the order parameter for the chiral phase transition. When the
current quark mass goes to infinity $m\rightarrow \infty$, QCD
becomes pure gauge $SU(3)$ theory, which is center symmetric in the
vacuum, and the usually used order parameter is the Polyakov loop
expectation value $\langle L \rangle $ \cite{Polyakov:1978vu}, which
is related to the heavy quark free energy.

In Sec.\ref{section-eos}, we have investigated thermodynamic
properties of two deformed ${\rm AdS}_5$ models with quadratic
corrections, and have found that both models agree well with lattice
results for pure $SU(3)$ gauge theory. Therefore, we will not
discuss the chiral phase transitions, but focus on deconfinement
phase transition properties of these two models. In this section, we
study the heavy quark potential, the Polyakov loop and the spatial
Wilson loop at finite temperature, which are quantities related to
deconfinement properties. It is worthy of mentioning that till now
there is no good method to calculate the properties of these loop
operators in the whole temperature region within the framework of
field theory, though chiral phase transition can be described by
using effective QCD models, e.g. the Nambu--Jona-Lasinio (NJL) model
\cite{NJL} and the linear sigma model \cite{LSM}. The gauge/gravity
duality offers a non-perturbative method to calculate the loop
operators, in return, lattice QCD results
\cite{Kaczmarek:2004gv,Gupta:2007ax,Bali:1993tz} of these loop
operators will judge the validity of the hQCD model.

At last, we want to mention that all the quantities of loop operator
will be calculated in the string frame, which is different from the
way of calculating the thermal quantities for equation of state. In
order to avoid confusion and also to keep this section
self-contained, we explicitly list the metric and the solved dual
black-hole background from Sec.\ref{sec-solution-hQCD}, which will
be used in this section:
\begin{eqnarray} \label{stringframe}
ds_S^2& = & \frac{L^2
e^{2A_s}}{z^2}\left(-f(z)dt^2+\frac{dz^2}{f(z)}+dx^{i}dx^{i}\right),
\nonumber \\
A_s &=& c k^2 z^2, \nonumber \\
f(z) & = & 1- f_{c} \int_0^{k z} x^3 \exp \left(\frac{3}{2} c x^2
\left( H_c(x/k)-1 \right)\right) dx,
\end{eqnarray}
with $c=\pm$, and $f_c$ and $H_c$ defined in Eq.(\ref{fc}) and
(\ref{Hc}), respectively.

\subsection{The heavy quark potential}
\label{section-HQ}

We firstly study the heavy quark potential. The linear heavy quark
anti-quark potential normally indicates the confinement of quarks in
the vacuum. Above the deconfinement critical temperature $T_c$, it
is expected that the linear potential vanishes and the Coulomb
potential to be exponentially screened at large distance. The heavy
quark potential at finite temperature has been analyzed in lattice
QCD \cite{Kaczmarek:2004gv}.

In the framework of gauge/gravity duality, we follow the standard
procedure \cite{Maldacena:1998im} and \cite{Rey:1998bq} to derive
the static heavy quark potential $V_{Q{\bar Q}}(r)$ under the
general metric background with a black hole, i.e, Eq.
(\ref{stringframe}). In $SU(N)$ gauge theory, the interaction
potential for infinity massive heavy quark antiquark is calculated
from the Wilson loop
\begin{equation}
W[C]=\frac{1}{N} Tr P \exp[i \oint_{C} A_\mu dx^\mu],
\label{Wilson-loop-formula}
\end{equation}
where $A_{\mu}$ is the gauge field, the trace is over the
fundamental representation, $P$ stands for path ordering. $C$
denotes a closed loop in space-time, which is a rectangle with one
direction along the time direction of length $T$ and the other space
direction of length $R$. The Wilson loop describes the creation of a
$Q{\bar Q}$ pair with distance $r$ at some time $t_0=0$ and the
annihilation of this pair at time $t=T$. For $T\to\infty$, the
expectation value of the Wilson loop behaves as $\langle
W(C)\rangle\propto e^{-T V_{Q\bar Q}}$. According to the
\textit{holographic} dictionary, the expectation value of the Wilson
loop in four dimensions should be equal to the string partition
function on the modified ${\rm AdS}_5$ space, with the string world
sheet ending on the contour $C$ at the boundary of ${\rm AdS}_5$
\begin{equation}
\langle W^{4d}[C]\rangle=Z_{string}^{5d}[C]\simeq e^{-S_{NG}[C]} \,\
,
\end{equation}
where $S_{NG}$ is the classical world sheet Nambu-Goto action
\begin{equation} \label{S-NG-HQ}
S_{NG}=\frac{1}{2\pi\alpha_q}\int d^2 \eta \sqrt{{\rm Det} \chi_{a
b}},
\end{equation}
with $\alpha_q$ the string tension which has dimension of ${\rm
GeV}^{-2}$, and $\chi_{ab}$ is the induced worldsheet metric with
$a,b$ the indices in the ($\eta^0=t,\eta^1=x)$ coordinates on the
worldsheet.

We consider the following situation there are static quark and
ant-quark linked by one string. The position of one quark is
$x=-\frac{r}{2}$ and the other is $x=\frac{r}{2}$. Under the
background (\ref{stringframe}), we can obtain the equation of
motion:
\begin{equation}\label{equa-HQP}
 \frac{\sqrt{f(z)}}{z^2} \frac{e^{2\mathcal
{A}_s(z)}}{\sqrt{1+\frac{(z')^2}{f(z)}}}= \text{Constant}=\sqrt{f_0}
\frac{e^{2\mathcal{A}_s(z_0)}}{z_0^2},
\end{equation}
Here the $r$ is dependent on $z_0$ which is the maximal value of $z$
and $ z'(x=0)=0$. In eq.(\ref{equa-HQP}), we have defined
$f_0=f(z=z_0)$. For the configuration mentioned above and the given
equation of motion, we impose the following boundary condtions $
z(x=0)=z_0,
 z(x=\pm \frac{r}{2})=0$. Following the standard procedure, one can
derive the interquark distance $r$ as a function of $z_0$
\begin{eqnarray}
r(z_0) &=& \int_0^1 d \nu \frac{2z_0 \nu^2}{\sqrt{1-\frac{f_0}{f(\nu
z_0)}\nu^4 \frac{e^{4A_s(z_0)}}{e^{4A_s(z_0
\nu)}}}}\frac{\sqrt{f_0}}{f(z_0
\nu)}\frac{e^{2A_s(z_0)}}{e^{2A_s(z_0 \nu)}}. \label{distance1}
\end{eqnarray}
The heavy quark potential can be worked out from the Nambu-Goto
string action:
\begin{eqnarray} V_{Q\bar
Q}(z_0)&=& \frac{g_q }{\pi z_0}\int_0^1 d \nu
\frac{1}{\sqrt{1-\frac{f_0}{f(\nu z_0)}\nu^4
\frac{e^{4A_s(z_0)}}{e^{4A_s(z_0
\nu)}}}}\frac{e^{2A_s(z_0\nu)}}{\nu^2}, \label{VQQ-general}
\end{eqnarray}
with $g_q=\frac{L^2}{\alpha_q}$. It is noticed that the integral in
Eq.(\ref{VQQ-general}) in principle include some poles, which
induces $ V_{Q\bar Q}(z)\rightarrow \infty$. The infinite energy
should be extracted through certain regularization procedure. The
divergence of $V_{Q\bar Q}(z)$ is related to the vacuum energy for
two static quarks. Generally speaking, the vacuum energy of two
static quarks will be different in various background. In our latter
calculations, we will use the regularized $V_{Q\bar Q}^{ren.}$ ,
which means the vacuum energy has been subtracted.

\begin{figure}[h]
\begin{center}
\epsfxsize=6.5 cm \epsfysize=6.5cm \epsfbox{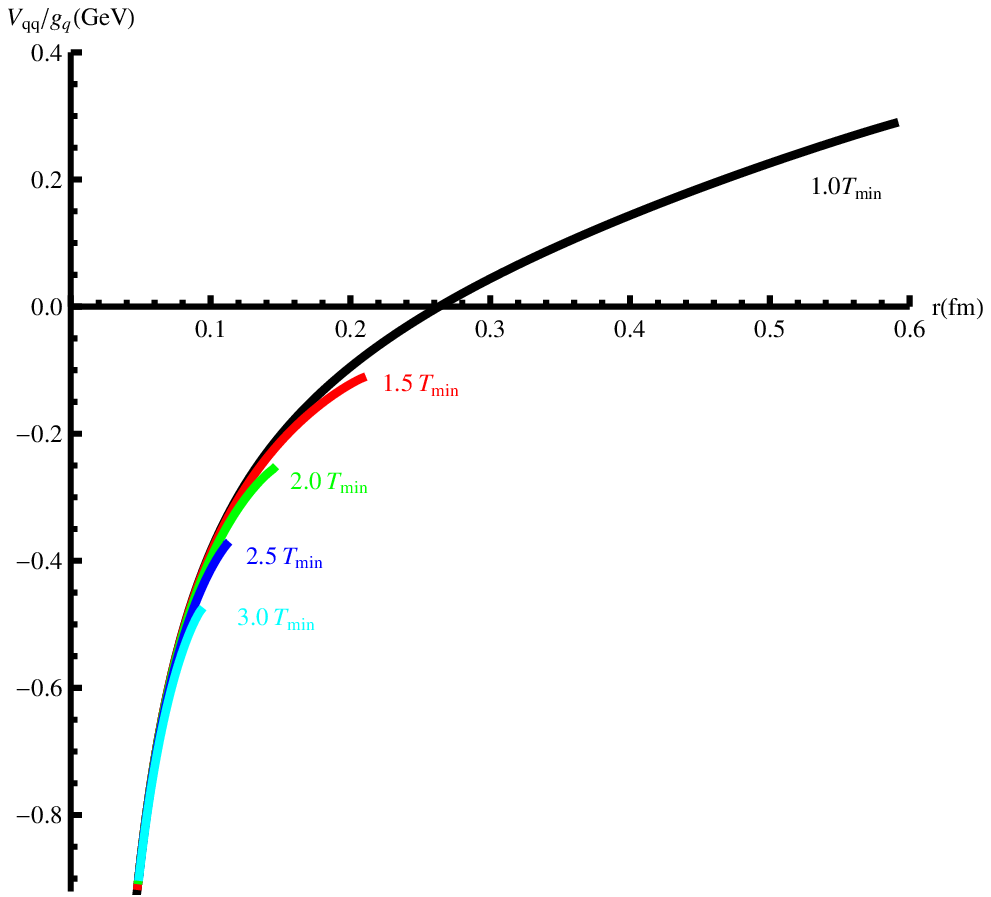}
\hspace*{0.1cm} \epsfxsize=6.5 cm \epsfysize=6.5 cm
\epsfbox{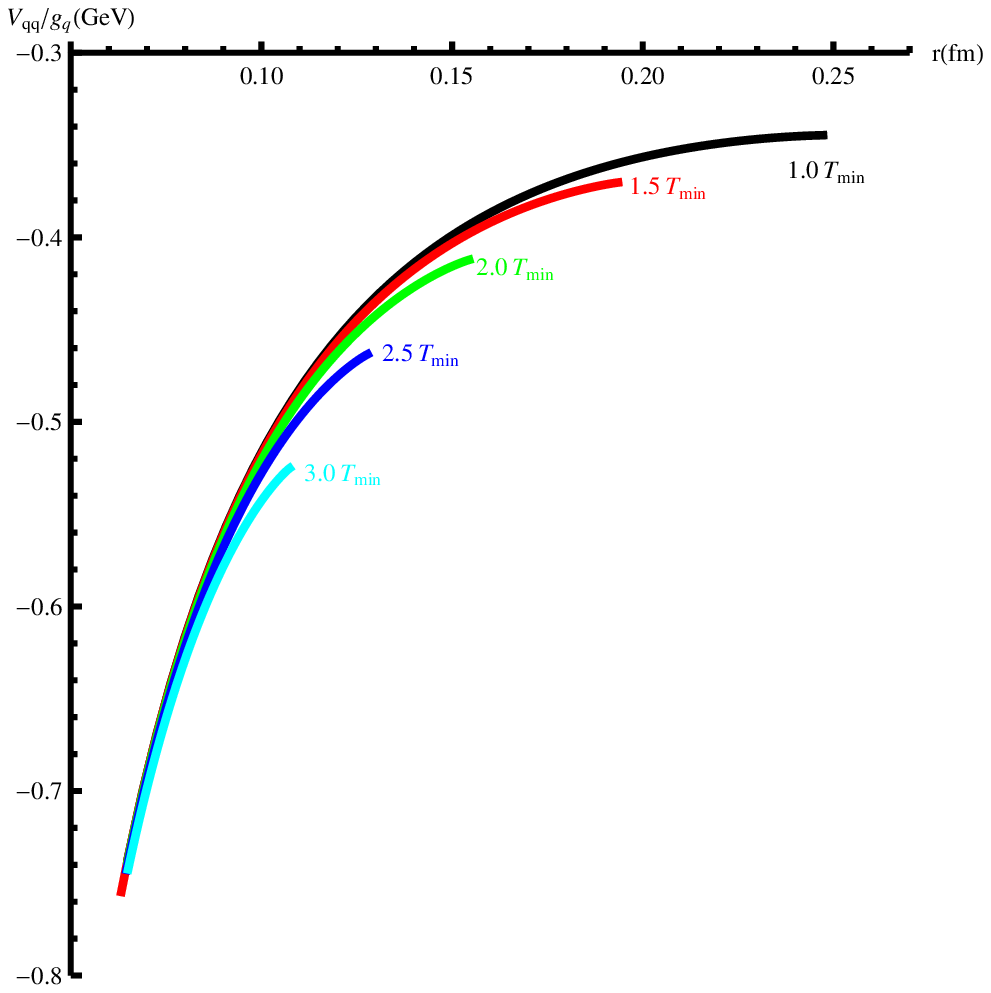} \vskip -0.05cm \hskip 0.15 cm
\textbf{( a ) } \hskip 6.5 cm \textbf{( b )} \\
\end{center}
\caption{The heavy quark potential $V_{Q\bar{Q}}/g_q$ as a function
of the separation distance $r$ for different temperatures for c=+
(a) and $c=-$ (b), respectively. Where $k=0.43 {\rm GeV}$ is used.}
 \label{HVQQ}
\end{figure}

Fig.\ref{HVQQ} shows the heavy quark potential $V_{Q\bar{Q}}/g_q$ as
a function of the distance between the two static quarks $r$ for
different temperatures $T/T_{min}=1,1.5,2,2.5,3$. Fig.\ref{HVQQ}(a)
is for $c=+$, i.e, the positive quadratic correction model, and
Fig.\ref{HVQQ}(b) is for $c=-$, i.e, the negative quadratic
correction model. For $c=+$, the heavy quark potential has a linear
part at zero temperature as shown in Ref. \cite{He:2010ye}. It is
observed that when $T> T_{min}$, the linear part vanishes, and with
the increases of the temperature, the Coulomb part potential gets
screened. This picture qualitatively agrees with lattice result in
Ref. \cite{Kaczmarek:2004gv}. For the case of $c=-$, the heavy quark
potential has no linear part at zero temperature as shown in Ref.
\cite{He:2010ye}. It is seen that when $T/T_{min} > 1$, the heavy
quark potential only contains Coulomb part, with the increase of the
temperature, the Coulomb part gets screened.

We can see that, even though the heavy quark potential for hQCD
models with positive and negative correction are quite different in
the vacuum, the behavior of the heavy quark potential in these two
hQCD models are qualitative the same above the critical temperature
$T_c$. The imaginary part of the heavy quark potential at finite
temperature in the framework of AdS/CFT has been discussed in Ref.
\cite{Im-HQ-Kovchegov}.

\subsection{The Polyakov loop}

The Polyakov loop is defined as
\begin{equation}\label{P1}
L(T)=\frac{1}{N}\text{tr Pexp}\Bigl[ig\int_0^{1/T}dt\, A_0 \Bigr]
\,,
\end{equation}
where $N$ is the colors, the trace is evaluated in the fundamental
representation and $P$ stands for the path ordering. In QCD with
infinitely heavy quarks, the Polyakov loop is related to the
operator that generates a static quark \cite{Polyakov:1978vu}. We
can interpret the logarithm of the expectation value
$\langle\,L(T)\,\rangle$ as half of the free energy $F_{Q{\bar Q}}$
of a static quark-anti-quark pair at infinite distance. The Polyakov
loop is an order parameter for center symmetry of the gauge group.
In the confined phase, the $F_{Q{\bar Q}}\rightarrow \infty$, which
ensures $\langle\,L(T)\,\rangle=0$, and thus the confined phase is
center symmetric. The deconfined phase is characterized by
$F_{Q{\bar Q}}< \infty$ and $\langle\,L(T)\,\rangle \neq 0$, which
implies the center symmetry is breaking in the ordered phase.

In the framework of gauge/gravity duality, the Polyakov loop is the
Wilson loop wrapping the periodic imaginary time direction
\cite{Witten:1998zw}. This operator should be computed in the string
frame \cite{Polyakov:1997tj}. We follow the method in Ref.
\cite{Andreev-T3} to calculate the expectation value of the Polyakov
loop, which is schematically given by the world-sheet path integral
\begin{equation}\label{pol}
\langle\,L(T)\,\rangle=\int DX\,e^{-S_w} \,,
\end{equation}
where $X$ denotes a set of world-sheet fields. $S_w$ is a
world-sheet action. In principle, the integral \eqref{pol} can be
evaluated approximately in terms of minimal surfaces that obey the
boundary conditions. The result is written as
$\langle\,L(T)\,\rangle=\sum_n w_n\exp[-S_n]$, where $S_n$ means a
renormalized minimal area whose weight is $w_n$.

Given the background in the string frame Eq.(\ref{stringframe}), we
can calculate the expectation value of the Polyakov loop by using
the Nambu-Goto action for $S_{w}$
\begin{equation} \label{S-NG-Polyakov}
  S_{NG} =  \frac{1}{2\pi
  \alpha_p}\int d^2\eta \sqrt{{\rm Det}\chi_{ab}},
\end{equation}
with $\alpha_p$ the string tension and $\chi_{ab}$ the induced
worldsheet metric with $a,b$ the indices in the
$(\eta^1=t,\eta^2=z)$ coordinates on the worldsheet. It is noticed
that here we have introduced a different string tension from that in
Eq.(\ref{S-NG-HQ}). This can be understood that along different
direction, the string tension is different. We will also introduce
another different string tension parameter for the spatial Wilson
loop.
This yields
\begin{equation}\label{2ng}
S_{\mathrm{NG}}=\frac{g_p }{\pi T}\int^{z_h}_0 dr\,\frac{
e^{2A_s}}{z^2} \sqrt{1+f(z) (\vec{x}\,')^2} \,,
\end{equation}
where $g_p=\frac{L^2}{2\alpha_p}$. A prime stands for a derivative
with respect to $z$.

\begin{figure}[h]
\begin{center}
\epsfxsize=6.5 cm \epsfysize=6.5 cm \epsfbox{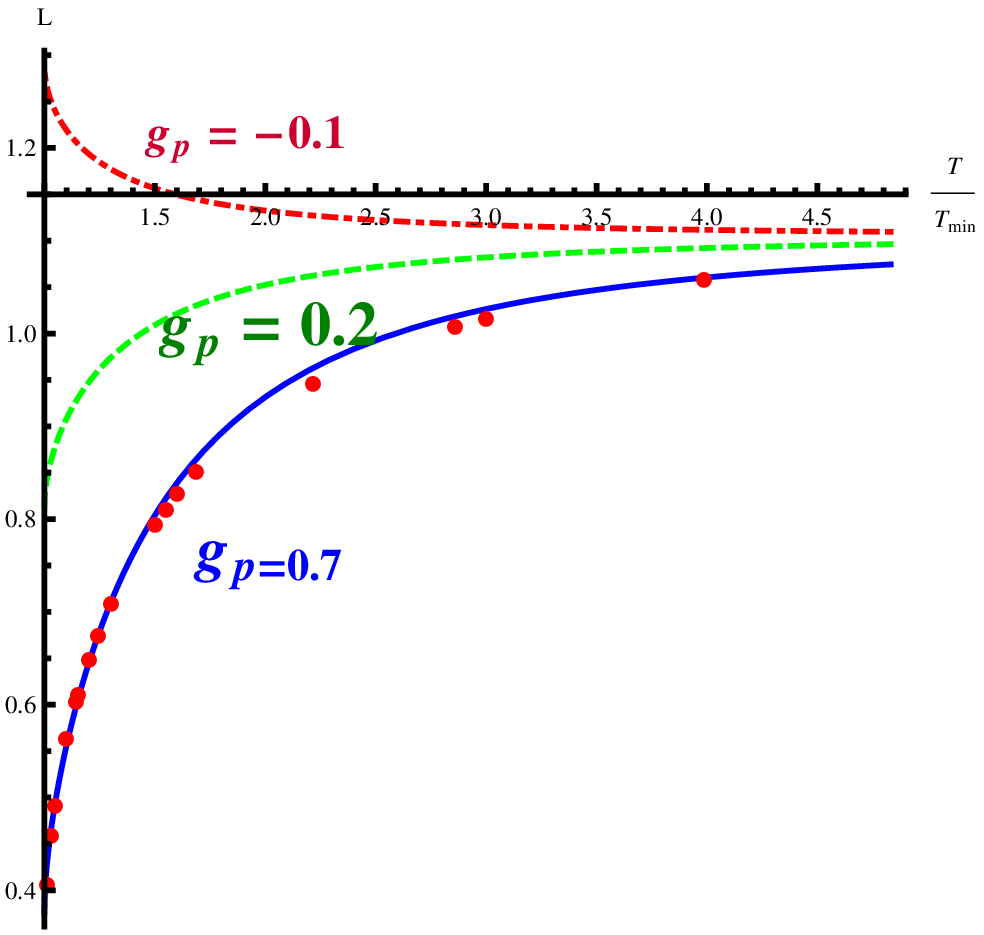}
\hspace*{0.1cm} \epsfxsize=6.5 cm \epsfysize=6.5 cm
\epsfbox{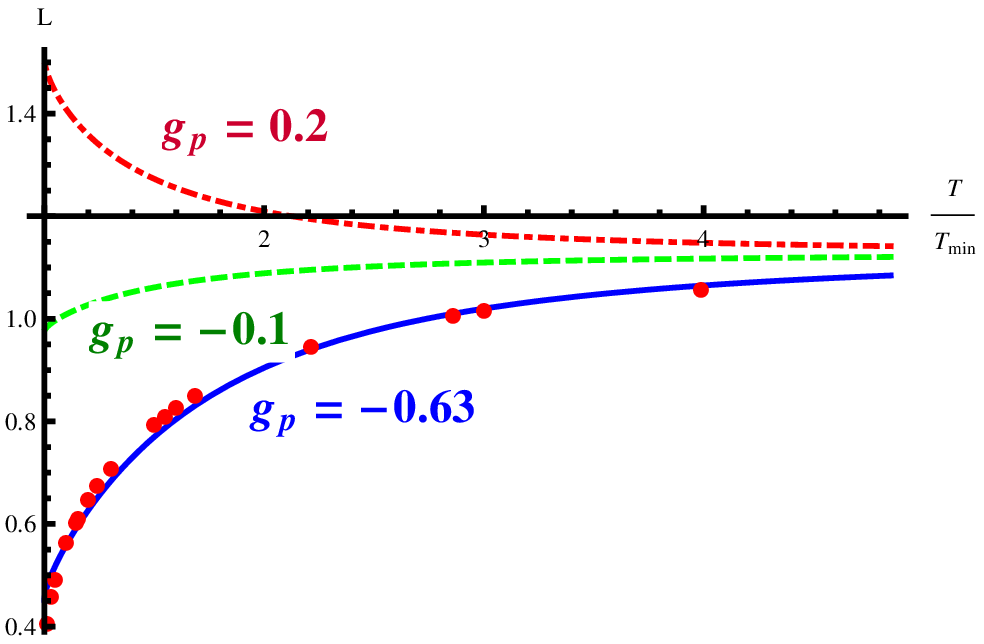} \vskip -0.05cm \hskip 0.15 cm
\textbf{( a ) } \hskip 6.5 cm \textbf{( b )} \\
\end{center}
\caption{The expectation value of the Polyakov loop as a function of
the temperature $T/T_{min}$: (a) in the case of $c=+$ with $k=0.43
GeV$, $C_p=0.1$ and $g_{p}=-0.1,0.2,0.7$, (b) in the case of $c=-$
with $k=0.43 GeV$, $C_p=0.12$, and $g_{p}=0.2,-0.1,-0.63$. The dots
are lattice data as a function of $T/T_c$ taken from
\cite{Gupta:2007ax}.} \label{Polyakovloop}
\end{figure}

The equation of motion for $\vec{x}$ takes the form of
\begin{equation}\label{eq-x}
\biggl[\frac{ e^{2A_s}}{z^2}
f(z)\vec{x}\,'/\sqrt{1+f(z)(\vec{x}\,')^2}\biggr]'=0 \,.
\end{equation}
The above equation has a special solution $\vec{x}=const$.
Substituting this constant solution to Eq.(\ref{2ng}), one can get
the minimal world-sheet,
\begin{eqnarray}\label{S0-Pol}
S_0 & = & S_0'+c_p, \nonumber \\
S_0' &= & \frac{g_p}{\pi T} \int_0^{z_h} d z \left(\frac{
e^{2A_s}}{z^2}- \frac{1}{z^2}\right).
\end{eqnarray}
Where $c_p$ is the normalization constant which is dependent on the
scheme of the regularization procedure. It is noticed that
Eq.(\ref{S0-Pol}) takes the same form as that in \cite{Andreev-T3},
where the black-hole background is taken from pure ${\rm AdS}_5$.
Therefore, Performing the integral over $z$, we get the same
analytic expression as in \cite{Andreev-T3}: \bea
S_0'&=&\frac{g_p}{\pi  T}\frac{ \sqrt{2 \pi } k z_h
Erfi\left(\sqrt{2} k z_h\right)-e^{2 k^2 z_h{}^2}+1}{z_h },\text{{ }}\text{{ }}\text{{ }}c=+,\\
S_0'&=&\frac{g_p}{\pi  T}\frac{ -\sqrt{2 \pi } k z_h
\text{Erf}\left(\sqrt{2} k z_h\right)-e^{-2 k^2 z_h^2}+1}{
z_h},\text{{ }}\text{{ }}\text{{ }}c=-. \eea Combining the weight
factor with the normalization constant as $C_p=\ln w_0-c_p$, we find
\begin{equation}\label{P2}
\langle L(T) \rangle = \exp \Big\{ C_p- S_0'\Big\}.
\end{equation}

The numerical result of the expectation value of the Polyakov loop
as a function of the scaled temperature $T/T_c$ is shown in Fig.
\ref{Polyakovloop} (a) and (b) for positive and negative quadratic
correction models, respectively. The dots are lattice data taken
from Ref.\cite{Gupta:2007ax}. For the positive quadratic correction
model, i.e, $c=+$, the parameters $k=0.43 GeV$ and $C_p=0.1$ have
been used for numerical calculation. We plot the Polyakov loop as a
function of temperature for several values of $g_p=-0.1,0.2,0.7$, it
is found that $g_p=0.7$ can remarkably fit the lattice data
\cite{Gupta:2007ax} for the pure $SU(3)$ gauge theory. For the
negative quadratic correction model, i.e, $c=-$, the parameters
$k=0.43 GeV$ and $C_p=0.12$ have been used for numerical
calculation. We plot the Polyakov loop as a function of the
temperature for several values of $g_p=0.2,-0.1,-0.63$, it is
observed that $g_p=-0.63$ can fit the lattice result quite well.

It can be seen that both models by choosing different parameters can
agree well with the lattice result on the Polyakov loop. However, we
have to point out that in the case of $c=-$, a negative value
$g_{p}=-0.63$ has been taken to fit the lattice data. This means the
string tension $\alpha_p$ introduced in Eq.(\ref{S-NG-Polyakov}) is
negative, which is not physical. Therefore, from the result of the
expectation value of the Polyakov loop, the negative quadratic
correction model can be excluded. Our result of the Polyakov loop
for positive quadratic correction model fits lattice data
remarkably.

\subsection{The spatial Wilson loop}
\label{sec-spatial-string}

We now move on to discuss the spatial Wilson loop in the framework
of the gauge/gravity duality.

The non-Abelian gauge theories undergo a deconfining phase
transition at high temperature. The physical string tension,
characterizing the linear rise of the potential between static quark
sources with distance, decreases with increasing temperature and
vanishes above $T_c$. The potential becomes a Debye screened Coulomb
potential in the high temperature phase, which is shown in
Sec.\ref{section-HQ}. However, the phase just above $T_c$ is more
complicated than a weakly coupled quark gluon gas because of the
appearance of some non-perturbative soft modes, e.g, the Debye
screening mass $m_D\sim g T$ and the magnetic screening mass $m_M
\sim g^2T$. The soft mode in magnetic sector cannot be handled in
any perturbative scheme. It is found that the non-perturbative
physics arising from the magnetic sector reflects the survival of
the area law behavior for space-like Wilson loops, i.e. confinement
of magnetic modes. The spatial string tension, $\sigma_s$, extracted
from the area law behavior of spatial Wilson loops, has been studied
in lattice calculations \cite{LAT-EOS-G,Bali:1993tz}, and shows the
expected dependence on the magnetic scale $\sqrt{\sigma_s}\sim
g^2(T) T$.

In the framework of gauge/gravity duality, to calculate the spatial
Wilson loop, we consider a rectangular loop $\mathcal{C}$ along two
spatial directions $(x, y)$ \cite{Andreev-T1}. The loop
$\mathcal{C}$ should wrap around the $\mathcal{S}^1$£¬ which is a
circle in the time direction. The expectation value of the loop can
be calculated through the following AdS/CFT dictionary:
\begin{equation}
\label{spatial-Wilson} V_s =\int D X e^{-S_w}.
\end{equation}
Here $X$ is the series of world sheet fields and $S_w$ stands for a
world sheet action. Again, at large N, the saddle point
approximation is valid, and we only need to take the minimum value
of the Euclidean action among the saddle points. In fact, given the
boundary conditions, there will be UV divergence from the field
theory viewpoint. It is necessary to regularize and make it finite
by a divergent subtraction by introducing a counterterm.

Given the background in the string frame Eq.(\ref{stringframe}), we
can calculate the expectation value of the spatial Wilson loop by
using the Nambu-Goto action for $S_{w}$
\begin{equation}
  S_{NG} =  \frac{1}{2\pi
  \alpha'}\int d^2\eta \sqrt{{\rm Det}\chi_{ab}},
\end{equation}
with $\alpha'$ the string tension and $\chi_{ab}$ the induced
worldsheet metric with $a,b$ the indices in the
$(\eta^1=x,\eta^2=y)$ coordinates on the worldsheet. We take one of
the spatial direction $Y$ goes to infinity. The quark and anti-quark
are set at $x_i=\pm r/2$, we can get the Nambu-Goto action as
\begin{equation}\label{ng1}
S=\frac{g_{sv} }{2\pi}Y\int^{\,\tfrac{r}{2}}_{-\tfrac{r}{2}}
dx\,\frac{h}{z^2} \sqrt{1+\frac{z'^2}{f(z)}} \,,
\end{equation}
here $r$ stands for the separation between the static quark pair and
$g_{sv}= \frac{L^2}{\alpha'}$. From the (\ref{ng1}), one can easily
get the equation of motion for $x$:
\begin{equation}\label{eqm}
z z''+\left(f(z)+z'^2\right)\left(2-z \partial_z\ln h\right)
-\frac{1}{2} z(z')^2\partial_z\ln f(z)=0.
\end{equation}
Following the standard procedure, one can obtain the simple
equation:
\begin{equation}\label{int}
\frac{h}{z^2\sqrt{1+\frac{z'^2}{f(z)}}}=\frac{h_0}{z_0^2} \,.
\end{equation}
Where $h(z)=e^{2A_s(z)}$ and $h_0=h(z_0) $. From the equation of
motion, one can see that
\begin{equation}
\label{Lr} r(z_0)=z_0 \int_0^1 \frac{2 \nu ^2 e^{-2 c k^2 \left(\nu
^2-1\right) z_0^2}}{\sqrt{f(z_0\nu ) \left(1-\nu ^4 e^{-4 c k^2
\left(\nu ^2-1\right) z_0^2}\right)}} \, d\nu,
\end{equation}
where $z_0=z(x=0)$ which defined by $z'(x=0)=0$ is the maximum value
of $z$. The free energy can be read:
\begin{equation}\label{SXY}
S_{xy}(z_0)=\frac{g_{sv}}{\pi z_0}\int_0^1 \frac{e^{2 c k^2 \nu ^2
z_0^2}}{\nu ^2 \sqrt{f(z_0\nu ) \left(1-\nu ^4 e^{-4 c k^2 \left(\nu
^2-1\right) z_0^2}\right)}} \, d\nu .
\end{equation}

After subtracting the UV divergence, one can get the area of the two
dimensional minimal surface given by the classical configuration of
the Nambu-Goto action Eq.(\ref{int}), and the regularized spatial
Wilson loop has the following form: \begin{equation}
V_s(z_0)=\frac{g_{sv}}{\pi z_0}\left(\int _0^1\frac{1}{\nu
^2}\left(\frac{e^{2 c k^2 \nu ^2 z_0^2}}{\sqrt{f(z_0\nu )
\left(1-\nu ^4 e^{-4 c k^2 \left(\nu ^2-1\right)
z_0^2}\right)}}-1\right)d\nu -1\right).
\end{equation}

\begin{figure}[h]
\begin{center}
\epsfxsize=6.5 cm \epsfysize=6.5cm \epsfbox{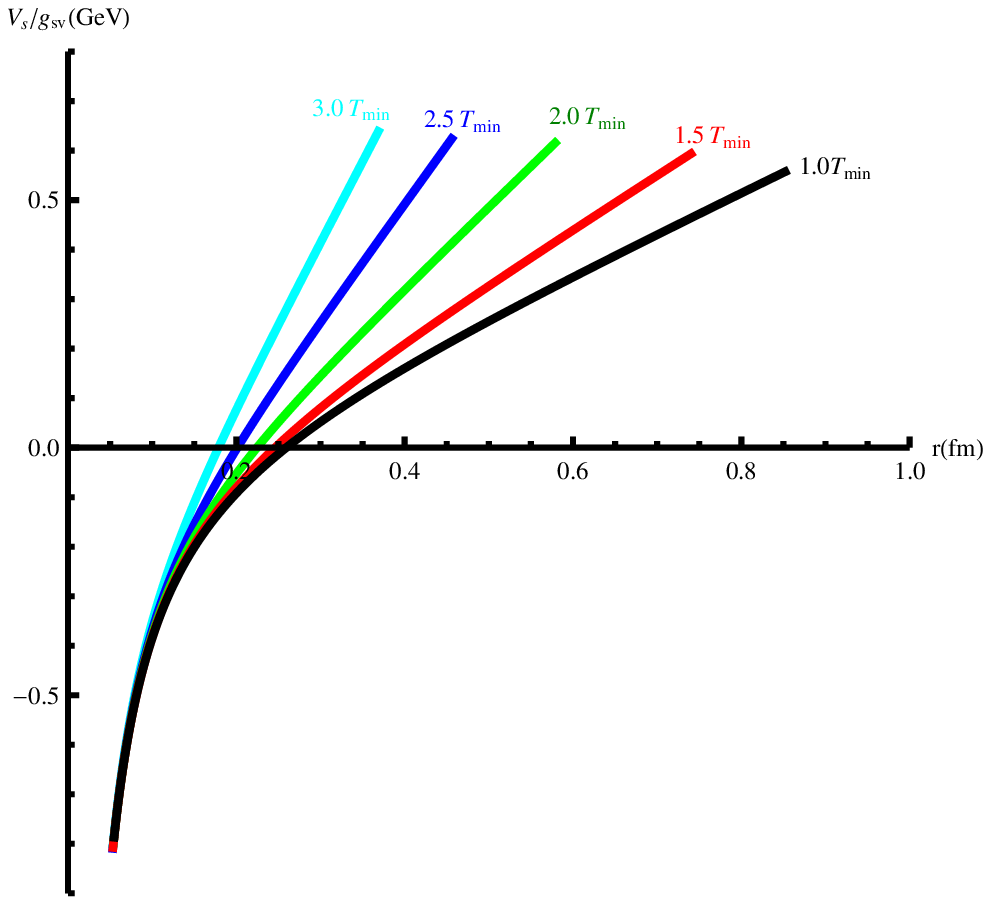}
\hspace*{0.1cm} \epsfxsize=6.5 cm \epsfysize=6.5 cm
\epsfbox{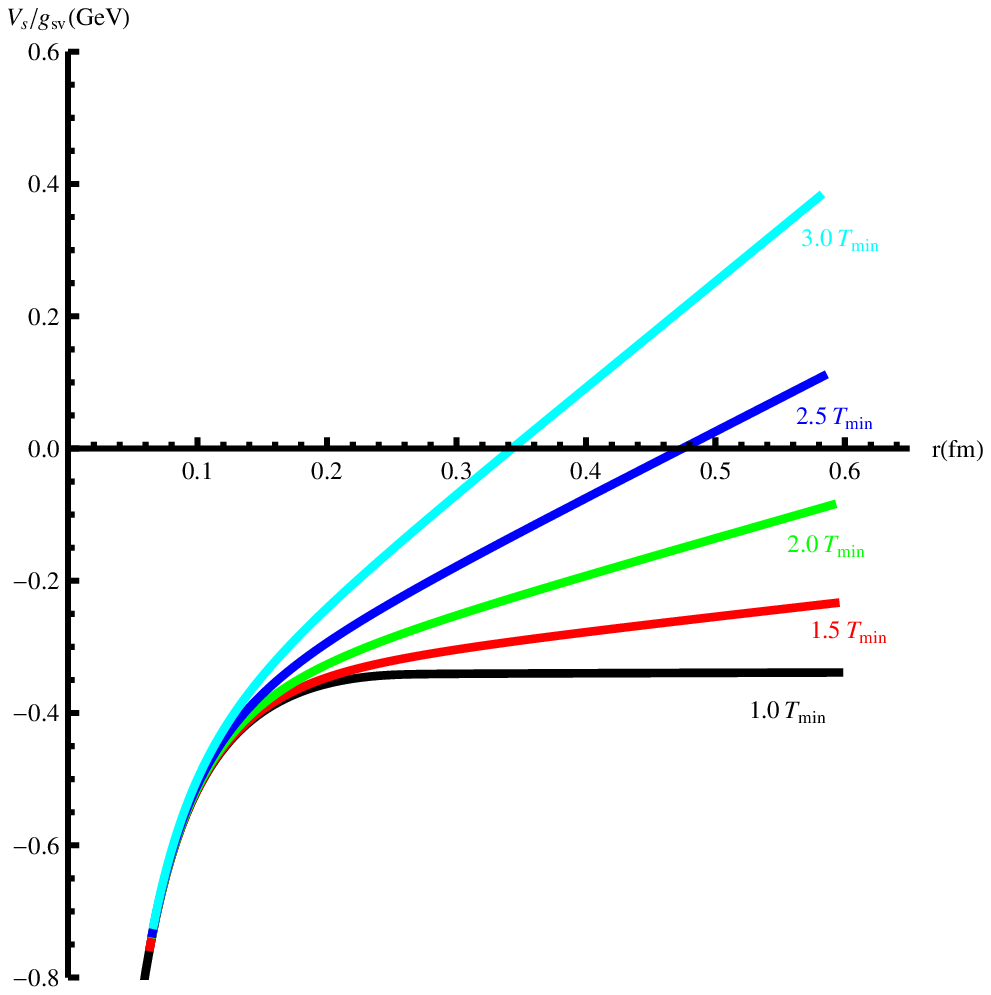} \vskip -0.05cm \hskip 0.15 cm
\textbf{( a ) } \hskip 6.5 cm \textbf{( b )} \\
\end{center}
\caption{The spatial heavy quark potential $V_s/g_{sv}$ as a
function the distance between the quark anti-quark $r$: (a) for the
case of $c=+$ with $k=0.43 {\rm GeV}$, (b) for the case of $c=-$
with $k=0.43 {\rm GeV}$. } \label{SXY-R}
\end{figure}

Fig. \ref{SXY-R} shows the regularized spatial potential $V_s$ as a
function of the distance between quark anti-quark $r$ for several
temperatures above $T_{min}$. For both $c=+$ and $c=-$ cases, the
parameter $k=0.43 {\rm GeV}$ is used. It is found that for positive
quadratic correction model, the spatial heavy quark potential always
have a linear potential, and the slope of the linear potential
increases with the temperature. This picture is in agreement with
lattice result \cite{LAT-EOS-G,Bali:1993tz}. For the case of $c=-$,
the spatial potential is flat at lower temperature, however, when
the temperature is high enough, we observe that a linear potential
rises up.

\begin{figure}[h]
\begin{center}
\epsfxsize=6.5 cm \epsfysize=6.5cm \epsfbox{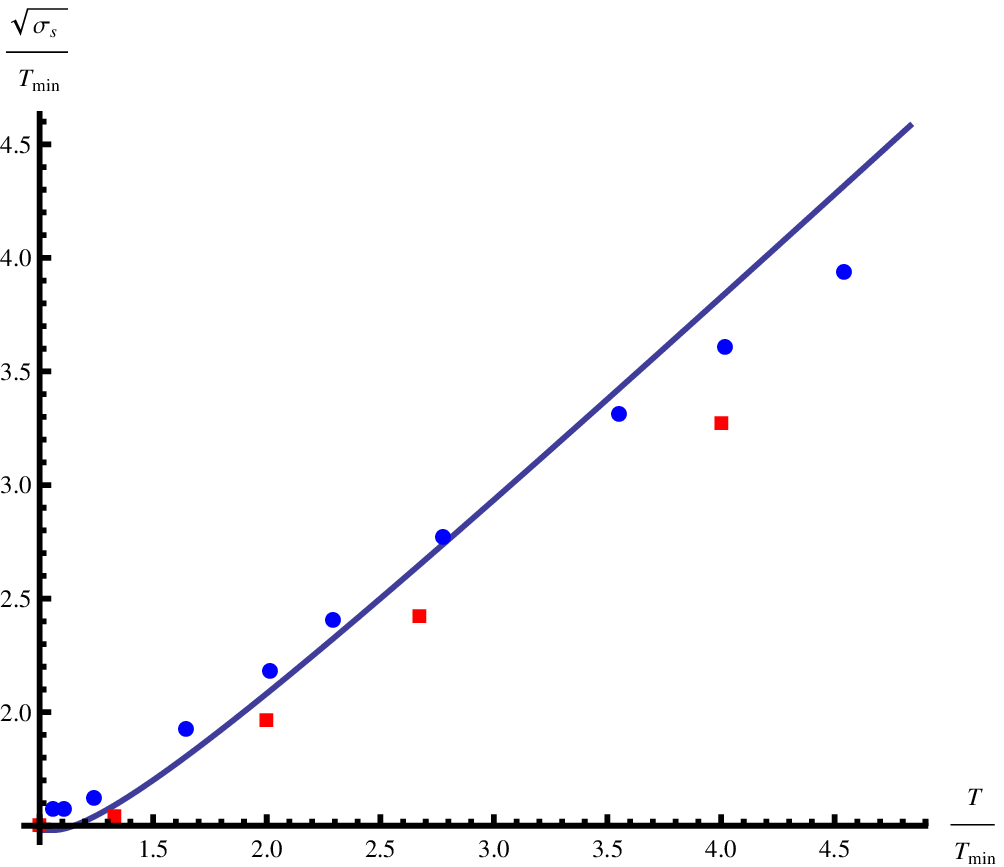}
\hspace*{0.1cm} \epsfxsize=6.5 cm \epsfysize=6.5 cm
\epsfbox{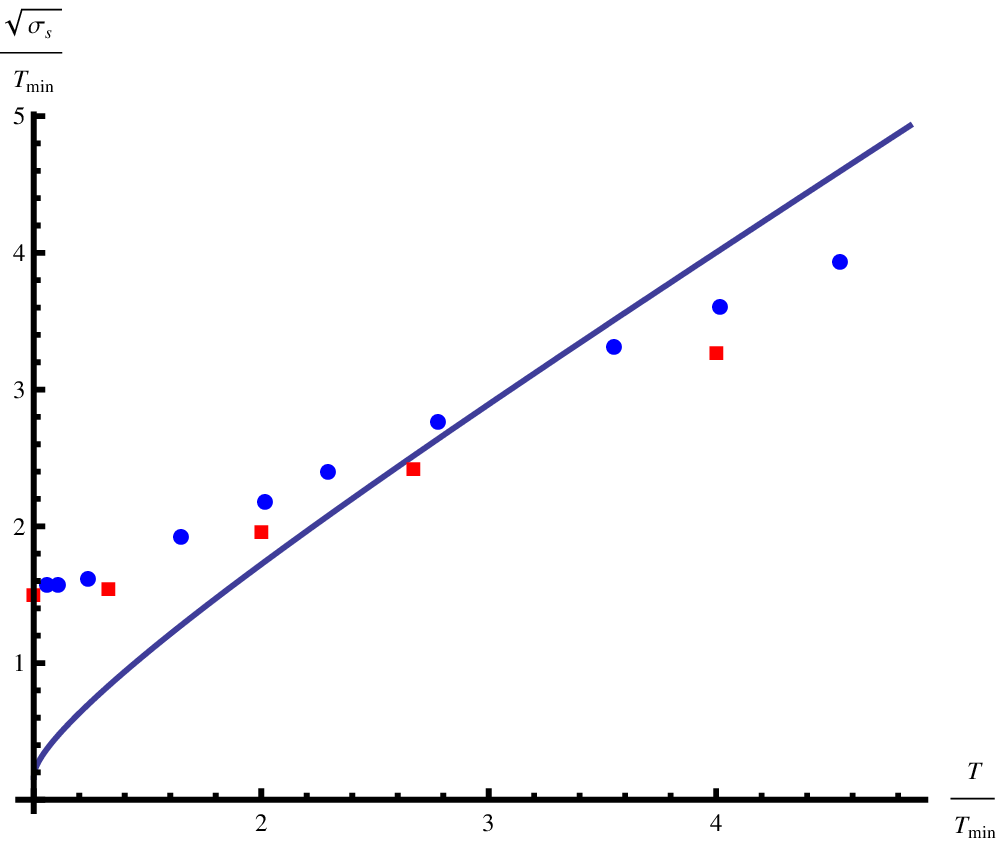} \vskip -0.05cm \hskip 0.15 cm
\textbf{( a ) } \hskip 6.5 cm \textbf{( b )} \\
\end{center}
\caption{The scaled spatial string tension $\sqrt{\sigma_s}/T_{min}$
as a function of the scaled temperature $T/T_{min}$ for the case of
$c=+$ in (a) with $k=0.43 {\rm GeV}$ and $g_{sv}=0.55$ are used, and
$c=-$ in (b) with $k=0.43 {\rm GeV}$ and $g_{sv}=0.7$ are used,
respectively. The blue dots stands for lattice data for pure $SU(3)$
gauge theory from \cite{ LAT-EOS-G}. The red dots are lattice data
as a function of $T/T_c$ for pure $SU(2)$ gauge theory which is from
\cite{Bali:1993tz}. } \label{spatialstringtension}
\end{figure}

Following the procedure introduced in Appendix
\ref{appendix-spatialtension}, we subtract the spatial string
tension $\sigma_s(T)$ and show the numerical results in Fig.
\ref{spatialstringtension}, where the lattice data for pure $SU(2)$
and $SU(3)$ gauge theory are taken from \cite{Bali:1993tz} and
\cite{LAT-EOS-G} respectively. For the case of $c=+$, it is observed
that the result of the spatial string tension fits very well with
the lattice data in Ref.\cite{LAT-EOS-G}\cite{Bali:1993tz}. While
for the model with negative quadratic correction, i.e. $c=-$, we
find that the spatial string tension is not in agreement with the
lattice date.

\section{Conclusion and discussion}
\label{section-summary}

From the previous studies of constructing the holographic QCD models
to describe hadron spectra and heavy quark potential, we have
observed that a quadratic correction is related to the confinement
property of QCD. There are two methods to introduce the quadratic
correction to hQCD models: the first method is to add it in the warp
factor of the metric, the second method is in the dilaton
background. However, the connection between these two methods is
lack of clarification. Moreover, the sign of the quadratic
correction is still under debate.

In this work, we have established a general framework for the
minimal graviton-dilaton system. By solving the Einstein equations,
we can self-consistently solve the dilaton background for any a
given metric structure assumed in a phenomenological holographic
model. Thus, the connection between the metric structure and the
dilaton background is clearly revealed. Furthermore, we can observe:
In the first method, the dilaton field should be solved out and its
contribution must be taken into account; In the second method, the
back-reaction of the dilaton background to the metric should be
considered. In other words, both methods to introduce the quadratic
correction are not self-consistent.

In this work, we extend the graviton-dilaton system to finite
temperature with a dual black-hole and self-consistently study the
sign of the quadratic correction in the deformed warp factor by
systematically investigating the thermodynamical properties and
comparing with lattice QCD on the results of the equation of state,
the heavy quark potential, the spatial Wilson loop and the Polyakov
loop. We find that the bulk thermodynamical properties are not
sensitive to the sign of the quadratic correction, and the results
of both deformed holographic QCD models with a positive and negative
quadratic correction agree well with lattice QCD results for pure
SU(3) gauge theory. However, the results from loop operator favor a
positive quadratic correction, which agree well with lattice
results. Especially, the result from the Polyakov loop excludes the
model with negative quadratic correction in the warp factor of ${\rm
AdS}_5$.

We would like to make one comment that the model with a quadratic
correction in the warp factor is not equivalent to the model with a
quadratic correction in the dilaton background. Therefore, our
result does not indicate the failure of the KKSS model, where the
negative quadratic correction is introduced in the dilaton
background. This deserves further study.

\begin{figure}[h]
\begin{center}
\epsfxsize=6.5 cm \epsfysize=6.5cm \epsfbox{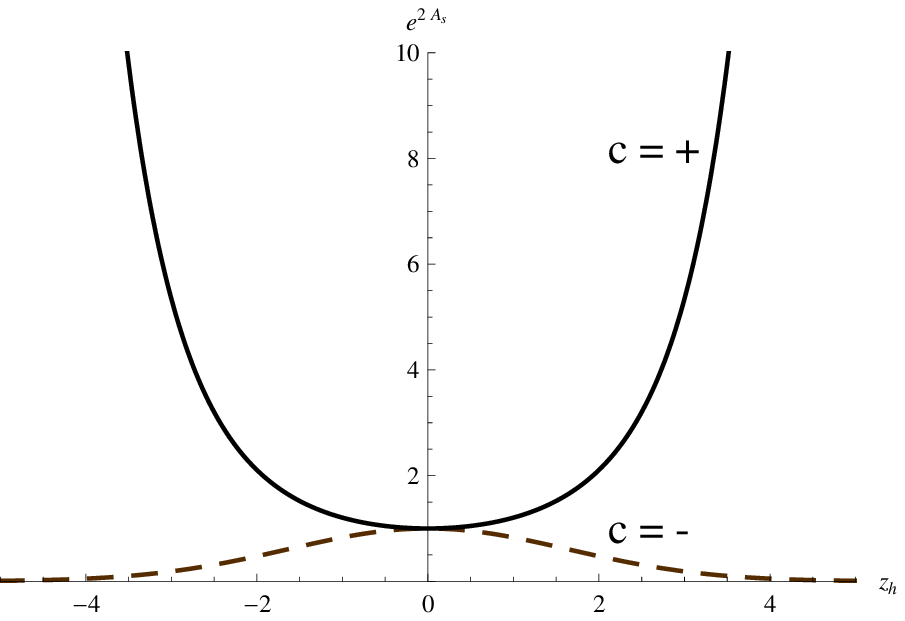} \hspace*{0.1cm}
\epsfxsize=6.5 cm \epsfysize=6.5 cm \epsfbox{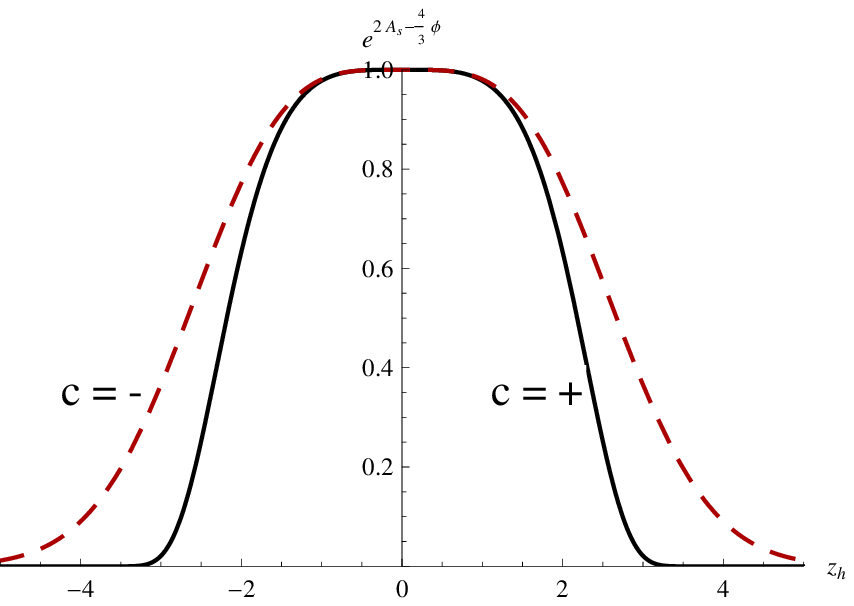} \vskip
-0.05cm \hskip 0.15 cm
\textbf{( a ) } \hskip 6.5 cm \textbf{( b )} \\
\end{center}
\caption{The warp factor in the string frame (a) and Einstein frame
(b) for $c=\pm$. } \label{twoframes}
\end{figure}

It is interesting to ask the question why the bulk thermodynamical
properties, such as the energy density, sound velocity, are not
sensitive to the sign of the quadratic correction in the deformed
warp factor, while the thermodynamical properties of loop operators
favor the positive quadratic correction model. The answer to this
question is offered below. The crucial point lies in that fact that
the bulk thermodynamical properties are defined in the Einstein
frame, and the thermal properties of loop operators are defined in
the string frame. In Fig.\ref{twoframes} (b), we can observe that
for positive and negative quadratic correction, the two warp factors
in the Einstein frame are almost the same. Therefore, the bulk
thermodynamical properties are not sensitive to the sign of the
quadratic correction. However, in Fig.\ref{twoframes} (a), we can
observe that for positive and negative quadratic correction, the two
warp factors in the string frame are quite sharp. This explains why
loop operators are sensitive to the sign of the quadratic
correction.

\vskip 1cm \noindent {\bf Acknowledgments}:

The authors thank O.Andreev, H. Liu, D.T. Son, Y. Tian, J.B. Wu,
J.F. Wu, X.N. Wu, W. Wang, and W.S. Xu and Y. Yang for valuable
discussions. S.H. thanks the support from the "TOP-100" project in
GUCAS. The work of M.H. is supported by CAS program "Outstanding
young scientists abroad brought-in", CAS key project KJCX2-EW-N01,
NSFC10735040, NSFC10875134, and K.C.Wong Education Foundation, Hong
Kong.

\appendix
\renewcommand{\theequation}{\thesection.\arabic{equation}}
\addcontentsline{toc}{section}{Appendices}
\section*{APPENDIX}

\section{Some simple analytical black hole solutions}
\label{appendix-solution}

In this appendix, we work out two analytical gravity solutions of
graviton-dilaton system by using Eq.(\ref{general-solution}) for
given metric ansatz in Einstein frame
Eq.(\ref{metric-Einsteinframe}). Here we are interested in the
solutions whose UV behavior is asymptotic $AdS_5$. We also impose
the constrains which are $f(0)=1$ near the $z\sim 0$, and requiring
$\phi(z), f(z)$ to be regular at $z=0, z_h$.

To avoid misunderstanding, we should clarify that we will list two
analytical gravity solutions in the Einstein frame. The metric
ansatz in the Einstein takes the following form:
\begin{eqnarray} \label{Ametric-Einsteinframe}
ds_E^2&=&\frac{L^2 e^{2A_s-\frac{4\phi}{3}}}{z^2}\left(-f(z)dt^2
+\frac{dz^2}{f(z)}+dx^{i}dx^{i}\right)\\
&=&\frac{L^2 e^{2A_E}}{z^2}\left(-f(z)dt^2
+\frac{dz^2}{f(z)}+dx^{i}dx^{i}\right).
\end{eqnarray}
Where $A_s(z)$ and $\phi(z)$ is defined by
Eq.(\ref{minimal-Einstein-action}) and Eq.
(\ref{metric-stringframe}) and $A_E(z)=A_s(z)-\frac{2\phi(z)}{3}$.

Using the general solutions in Eq.(\ref{general-solution}), one can
easily find some exact new gravity solutions. Here, we do not repeat
the procedure and details of the calculation to obtain the new
gravity solutions, we just list two simple analytical gravity
solutions for example.

The first solution is given as following:
\begin{eqnarray}
A_E(z)&=&\log \left(\frac{z }{z_0\sinh(\frac{z}{z_0})}\right),\nonumber\\
f(z)&=&1-\frac{V_{1}}{3}(2-3\cosh(\frac{z}{z_0})+\cosh^3(\frac{z}{z_0})),\nonumber\\
\phi(z)&=&\frac{3 z}{2 z_0}.
\end{eqnarray}
Where $z_0$ is integral constants and $V_{1}$ is the constant from
the dilaton potential \bea\label{dilatonpotential1}
V_E(\phi)&=&-\frac{16V_{1}(\sinh^6(\frac{\phi}{3}))+9(\sinh^2(\frac{2\phi}{3}))+12}{L^2}.
\eea The solution above is given by the graviton-dilaton system
(\ref{minimal-Einstein-action}) with nontrivial dilaton potential
$V_E$. Where the subscript $E$ stands for the dilaton potential in
the Einstein frame. Here we should emphasize that the exact relation
of the dilaton potential between the Einstein frame and string frame
is given by the Eq. (\ref{string-enssteinframe}).

The second new gravity solution is as below: \bea
A_E(z)&=&-\log \left(1+\frac{z}{z_0}\right),\\
f(z)&=& 1-V_{2} \left(\frac{z^7}{7 z_0^7}+\frac{z^6}{2 z_0^6}
+\frac{3 z^5}{5 z_0^5}+\frac{z^4}{4 z_0^4}\right),\\
\phi(z)&=&3 \sqrt{2} \sinh
^{-1}\left(\sqrt{\frac{z}{z_0}}\right).\eea Where $z_0$ is integral
constants and $V_{2}$ are the constants from the dilaton potential
\bea\label{dilatonpotential2} V_E(\phi)&=&-\frac{3 V_{2} \sinh
^{14}\left(\frac{\phi }{3 \sqrt{2}}\right)}{35 L^2}-\frac{3 V_{2}
\sinh ^{12}\left(\frac{\phi }{3 \sqrt{2}}\right)}{10
L^2}-\frac{3V_{2} \sinh ^{10}\left(\frac{\phi }{3
\sqrt{2}}\right)}{10
L^2}\nonumber\\
&{}&-\frac{42 \sinh ^4\left(\frac{\phi }{3
\sqrt{2}}\right)}{L^2}-\frac{42 \sinh ^2\left(\frac{\phi }{3
\sqrt{2}}\right)}{L^2}-\frac{12}{L^2}. \eea

In this context, we do not discuss the application of the above two
gravity solutions, such as the thermal dynamical quantities and
stability of the two black hole solutions, which are left to the
future work.

\section{Free energy density at UV boundary for the hQCD model with $A_s(z)=c k^2z^2$ }
\label{appendix-epsilon}

We offer the details to derive the free-energy density at UV
boundary $z=\epsilon\rightarrow 0$ for the hQCD model with
$A_s(z)=ck^2z^2$ described in Section \ref{sec-solution-hQCD}.

For black-hole case, the solutions in Section
\ref{sec-solution-hQCD} expanded around UV boundary have the
following expressions:
\begin{eqnarray}
\phi (z)=&&\frac{3c}{2}  k^2 z^2+\frac{3}{10}  k^4 z^4+\frac{4c}{35}
 k^6 z^6+\frac{4}{105}  k^8 z^8+o(z^9),\\
f(z)=&&1-\frac{f_c^h}{4}k^4z^4-\frac{3 f_c^h}{40}k^8z^8+o(z^9),\\
A_E(z)=&&-\frac{1}{5} k^4z^4 -\frac{8c}{105} k^6z^6 -\frac{8
}{315}k^8z^8+o(z^9).
\label{BH-epsilon}
\end{eqnarray}

For the case of thermal gas, the solutions at UV boundary take the
expression as:
\begin{eqnarray}
\phi_0(z)=&&p_2k^2z^2+\frac{2p_2^2}{15} k^4  z^4+\frac{
p_2^3 (128-105 f_c^h) }{3780}k^6z^6 \nonumber\\
&&+\frac{ p_2^4 (64-63 f_c^h)}{8505} k^8z^8+o(z^9),\\
f_0(z)=&&1,\\
A_{E0}(z)=&&-\frac{4 p_2^2}{45} k^4 z^4 -\frac{64p_2^3}{2835}
 k^6 z^6+ \frac{p_2^4 (-128 + 105 f_c^h)}{25515}
  k^8z^8+o(z^9).
  \label{TG-epsilon}
\end{eqnarray}

To compare the free-energy density of the black-hole case and the
thermal gas case, one has to impose the following conditions
\cite{H-H}:
\begin{eqnarray}
\phi_0(z)\mid_{\epsilon}=&& \phi(z)\mid_{\epsilon},\\
\tilde{\beta} b_0(z)\mid_{\epsilon}=&&\beta b(z)\mid_{\epsilon},\\
\tilde{V_3} b_0^3(z)\mid_{\epsilon}=&&V_3 b^3(z)\mid_{\epsilon}.
\end{eqnarray}
Comparing Eq.(\ref{BH-epsilon}) with Eq.(\ref{TG-epsilon}), we can
get the relationship between $p_2$ and $k$,
\begin{eqnarray}
{p}_2=&&\frac{3}{2} c \left(1+\frac{1}{16} k^4 f_c^h \epsilon
^4+\frac{k^8 \left(-128 f_c^h+105(f_c^h)^2\right) \epsilon
^8}{8960}+o(\epsilon^9)\right).
\end{eqnarray}
Then we get the free energy densities for the black-hole and thermal
gas at UV boundary as follows:
\begin{eqnarray}
{\cal F}_{BH}^{\epsilon}=&&2 M^3 \left(3 b(\epsilon )^2 f(\epsilon )
b'(\epsilon )+\frac{1}{2} b(\epsilon )^3 f'(\epsilon
)\right)\nonumber\\
=&&-\frac{6 \left(L^3 M^3\right)}{\epsilon ^4}-\left(\frac{6}{5} k^4
L^3 M^3-\frac{1}{2} k^4 L^3 M^3
f_c^h\right)+o(\epsilon^2)\\
{\cal F}_{TG}^{\epsilon}=&&\frac{2 M^3 \left(b(\epsilon )^4
\sqrt{f(\epsilon )}\right) \left(3 b_0(\epsilon )^2 f_0(\epsilon )
b_0'(\epsilon )+\frac{1}{2} b_0(\epsilon )^3 f_0'(\epsilon
)\right)}{b_0(\epsilon )^4
\sqrt{f_0(\epsilon )}}\nonumber\\
=&&-\frac{6 L^3 M^3}{\epsilon^4}-\left(\frac{6}{5} k^4 L^3
M^3-\frac{3}{4} k^4 L^3 M^3 f_c^h\right)+o(\epsilon^2).
\end{eqnarray}

From the above result,one can derive the free energy density
difference as follows:
\begin{eqnarray}
\Delta\mathcal{F}=-\frac{1}{4}k^4M^3L^3f_c^h-2M^3b_0^2(z_{IR})b_0^\prime(z_{IR}).
\end{eqnarray}
From Eqs.(\ref{fc-zh}),(\ref{temp}) and (\ref{entrpy}), one can
express the temperature dependent coefficient $f_c^h$ with the
entropy density and temperature as
\begin{eqnarray}
f_c^h=\frac{16\pi G_5}{k^4L^3 }Ts=\frac{Ts}{k^4M^3L^3}.
\end{eqnarray}
As a result, one gets
\begin{eqnarray}
\Delta\mathcal{F}=-\frac{1}{4}Ts-2M^3b_0^2(z_{IR})b_0^\prime(z_{IR}).
\label{dF-s}
\end{eqnarray}

\section{To extract the spatial string tension}
\label{appendix-spatialtension}

In this appendix, we show how to extract the spatial string tension.
A fist-principle method for determining the spatial string tension
in hot QCD matter is to measure the rectangular spatial Wilson loops
as we show in Sec.\ref{sec-spatial-string}.  The spatial potential
$V_s(r)$ is defined by:
\begin{equation}
V_s(r)=- \lim_{Y\rightarrow\infty} \frac{1}{r} \log W(r,Y).
\end{equation}
The spatial string tension can be obtained: \begin{equation}
\sigma_s (T) = \lim_{r\rightarrow \infty} \frac{V_s(r)}{r}.
\end{equation}
To extract the spatial string tension, one can use the following
relation: \bea V_s=S_{xy}\sim \sigma_s r \text{ { }{} } r \gg 1.
\eea Under the general background
\begin{equation}
ds^2=\frac{L^2
e^{2A_s}}{z^2}\left(f(z)dt^2+\frac{dz^2}{f(z)}+dx^{i}dx^{i}\right),
\end{equation}
the separation of the two static quarks has the form of:
\begin{equation}\label{RRR}
r(z_0)=z_0 \int_0^1 \frac{2 \nu ^2 e^{2 A_s\left(z_0\right)-2
A_s\left(\nu  z_0\right)}}{\sqrt{f\left(\nu  z_0\right) \left(1-\nu
^4 e^{4 A_s\left(z_0\right)-4 A_s\left(\nu  z_0\right)}\right)}}
\,d\nu .
\end{equation}
The free energy of the spatial Wilson loop is given by:
\begin{equation}\label{VVV}
S_{xy}(z_0)=\frac{g_{sv}}{\pi  z_0}\int _0^1\frac{1}{\nu
^2}\frac{e^{2 A_s\left(\nu z_0\right)}}{\sqrt{f\left(\nu z_0\right)
\left(1-\nu ^4 e^{4 A_s\left(z_0\right)-4 A_s\left(\nu
z_0\right)}\right)}}d\nu.
\end{equation}
We assume that the function $A_s(z_0)$ is a regular function here.
One should focus on the denominator of the Eq.(\ref{RRR}) and
Eq.(\ref{VVV}) which make main contribution to the integral when
$\nu \sim 1$. If one just focus on the dominant part, one can expand
the denominators at $\nu = 1$ respectively. After the expansion up
to the order $O((\nu-1)^2)$ and integral out the formula, one can
find the coefficient:
\begin{eqnarray} \label{coeff}
& &
(1-\nu)\left(-4 z_0 f\left(z_0\right) A_s'\left(z_0\right)+4
f\left(z_0\right)\right)\nonumber \\
& & +(1-\nu )^2 \Big(4 z_0^2 f'\left(z_0\right) A_s'\left(z_0\right)
 -8 z_0^2 f\left(z_0\right) A_s'\left(z_0\right){}^2+24 z_0 f\left(z_0\right)
 A_s'\left(z_0\right)\nonumber \\
 & & +2 z_0^2 f\left(z_0\right) A_s''\left(z_0\right)
 -4 z_0 f'\left(z_0\right)-14 f\left(z_0\right)\Big)
\end{eqnarray}
In term of the following useful integral formula:
 \begin{equation}\label{integral}
\int_0^1 \frac{1}{\sqrt{a (1-\nu)+b(1-\nu)^2}}=\frac{2
\text{ArcTanh}\left(\sqrt{\frac{b}{a+b}}\right)}{\sqrt{b}},\text{ {
}}\text{ }a>0 \text{ { }} \text{and } b>0,
\end{equation}
one can notice that the integral will be divergent when $a=0$.
$r(z_0)$ can be obtained and has the form of
\begin{equation}\label{leadr} r(z_0)=\frac{4 z_0 \tanh
^{-1}\left(\frac{\sqrt{b_L}}{\sqrt{a_L+b_L}}\right)}{\sqrt{b_L}}+O(1)
\end{equation}
with $a_L$ and $b_L$ the coefficients appeared in
Eq.(\ref{integral}) respectively.

Following the same strategy, we can get the spatial potential
\begin{equation}\label{leadV}
V_s(z_0)=\frac{2 e^{2A_s(z_0)} \tanh
^{-1}\left(\frac{\sqrt{b_V}}{\sqrt{a_V+b_V}}\right)}{\pi z_0
\sqrt{b_V}}+O(1). \end{equation} Where $a_V$ and $b_V$ are the
coefficients mentioned in Eq.(\ref{integral}) respectively.
Comparing Eq.(\ref{leadr}) and Eq.(\ref{leadV}), one can easily get
the spatial string tension. Up to the order $O(1-\nu)^2$, the
coefficients listed in Eq.\ref{coeff} have the following exact
relations such as $a_L= a_V$ and $b_L=b_V$ in Eq.(\ref{leadr}) and
Eq.(\ref{leadV}). The spatial string tension has the form of
\begin{equation}\label{spatialtension}
\sigma_s/g_{sv} = \frac{e^{2 A_s(z_h)}}{2\pi z_h^2},
\end{equation} Here,
One can find that the spatial string tension $\sigma_s$ depends on
the warp factor $A_s(z)$. In pure $AdS_5$ case, \bea\sigma_s/g_{sv}
= \frac{1}{2\pi z_h^2}. \eea

\end{document}